\documentclass[longauth]{aa}  

\usepackage{graphicx}
\usepackage{subcaption}
\usepackage[colorlinks=true,linkcolor=blue,citecolor=blue, urlcolor=blue]{hyperref}
\usepackage{tikz}
\usepackage{amsmath}
\usepackage{comment}
\usepackage{float}
\usepackage{soul}
\usepackage{tabularx,booktabs,makecell}
\usepackage{txfonts}
\usepackage{array}
\newcolumntype{P}[1]{>{\centering\arraybackslash}p{#1}}

\makeatletter
\renewcommand*\aa@pageof{, page \thepage{} of \pageref*{LastPage}}
\makeatother

\definecolor{RoyalBlue}{cmyk}{1, 0.50, 0, 0}

\begin{document} 

\title{$S^5$: New insights from deep spectroscopic observations of the tidal tails of the globular clusters NGC~1261 and NGC~1904}
\titlerunning{The extra-tidal features of globular clusters NGC 1261 and NGC 1904}
\author{
Petra~Awad\inst{\ref{kapteyn}, \ref{bernoulli}}, 
Ting S. Li\inst{\ref{toronto}}, 
Denis Erkal\inst{\ref{surrey}}, 
Reynier F. Peletier\inst{\ref{kapteyn}}, 
Kerstin Bunte\inst{\ref{bernoulli}}, 
Sergey E. Koposov\inst{\ref{edinbrugh}, \ref{cambridge1}, \ref{cambridge2}},
Andrew Li\inst{\ref{toronto}}, 
Eduardo Balbinot\inst{\ref{kapteyn}}, 
Rory Smith\inst{\ref{santiago}}, 
Marco Canducci\inst{\ref{birmingham}}, 
Peter Ti{\v{n}}o\inst{\ref{birmingham}}, 
Alexandra M. Senkevich\inst{\ref{surrey}},
Lara~R.~Cullinane\inst{\ref{leibniz}}, 
Gary~S.~Da~Costa\inst{\ref{canberra}, \ref{astro3D}},
Alexander~P.~Ji\inst{\ref{chicago1}, \ref{chicago2}}, 
Kyler~Kuehn\inst{\ref{lowell}},
Geraint~F.~Lewis\inst{\ref{sydney}},
Andrew~B.~Pace\inst{\ref{mcwilliams}}, 
Daniel~B.~Zucker\inst{\ref{macquarie1}, \ref{macquarie2}}, 
Joss~Bland-Hawthorn\inst{\ref{sydney}, \ref{astro3D}},
Guilherme~Limberg\inst{\ref{saopaulo}, \ref{chicago2}}, 
Sarah~L.~Martell\inst{\ref{unsw}, \ref{astro3D}},
Madeleine~McKenzie\inst{\ref{canberra}, \ref{astro3D}}, 
Yong~Yang\inst{\ref{sydney}}, 
Sam~A.~Usman\inst{\ref{chicago2}}
\\
($S^5$ Collaboration)
}

\authorrunning{Awad et al. }

\institute{
Kapteyn Astronomical Institute, University of Groningen, PO Box 800, 9700 AV Groningen, The Netherlands
\label{kapteyn}
\newline
\email{p.awad@rug.nl}
\and
Bernoulli Institute for Mathematics, Computer Science and Artificial Intelligence, University of Groningen, 9700AK Groningen, The Netherlands
\label{bernoulli}
\and
Department of Astronomy and Astrophysics, University of Toronto, 50 St. George Street, Toronto ON, M5S 3H4, Canada
\label{toronto}
\and
Department of Physics, University of Surrey, Guildford GU2 7XH, UK
\label{surrey}
\and
Universidad Technica Frederico de Santa Maria, Avenida Vicuña Mackenna 3939, San Joaquín, Santiago, Chile
\label{santiago}
\and
University of Birmingham, School of Computer Science, B15 1TT, Birmingham, United Kingdom
\label{birmingham}
\and
Institute for Astronomy, University of Edinburgh, Royal Observatory, Blackford Hill, Edinburgh EH9 3HJ, UK
\label{edinbrugh}
\and
Institute of Astronomy, University of Cambridge, Madingley Road, Cambridge CB3 0HA, UK
\label{cambridge1}
\and
Kavli Institute for Cosmology, University of Cambridge, Madingley Road, Cambridge CB3 0HA, UK
\label{cambridge2}
\and 
Leibniz-Institut f{\"u}r Astrophysik Potsdam (AIP), An der Sternwarte 16, D-14482 Potsdam, Germany
\label{leibniz}
\and 
Research School of Astronomy and Astrophysics, Australian National University, Canberra, ACT 2611, Australia
\label{canberra}
\and
Centre of Excellence for All-Sky Astrophysics in Three Dimensions (ASTRO 3D), Australia
\label{astro3D}
\and 
Department of Astronomy \& Astrophysics, University of Chicago, 5640 S Ellis Avenue, Chicago, IL 60637, USA
\label{chicago1}
\and
Kavli Institute for Cosmological Physics, University of Chicago, Chicago, IL 60637, USA
\label{chicago2}
\and
Lowell Observatory, 1400 W Mars Hill Rd, Flagstaff,  AZ 86001, USA
\label{lowell}
\and 
Sydney Institute for Astronomy, School of Physics, A28, The University of Sydney, NSW 2006, Australia
\label{sydney}
\and
McWilliams Center for Cosmology, Carnegie Mellon University, 5000 Forbes Ave, Pittsburgh, PA 15213, USA
\label{mcwilliams}
\and
School of Mathematical and Physical Sciences, Macquarie University, Sydney, NSW 2109, Australia
\label{macquarie1}
\and 
Macquarie University Research Centre for Astrophysics and Space Technologies, Sydney, NSW 2109, Australia
\label{macquarie2}
\and 
Universidade de S\~ao Paulo, Instituto de Astronomia, Geof\'isica e Ci\^encias Atmosf\'ericas, Departamento de Astronomia, SP 05508-090, S\~ao Paulo, Brasil
\label{saopaulo}
\and 
School of Physics, University of New South Wales, Sydney NSW 2052, Australia\\
\label{unsw}
}

   \date{Received \today; accepted }

 
\abstract{
As globular clusters (GCs) orbit the Milky Way, their stars are tidally stripped forming tidal tails that follow the orbit of the cluster around the Galaxy. The morphology of these tails is complex and shows correlations with the phase of orbit and the orbital angular velocity, especially for GCs on eccentric orbits. Here, we focus on two GCs, NGC~1261 and NGC~1904, that have potentially been accreted alongside Gaia-Enceladus and that have shown signatures of having, in addition to tidal tails, structures formed by distributions of extra-tidal stars that are misaligned with the general direction of the clusters' respective orbits. To provide an explanation for the formation of these structures, we make use of spectroscopic measurements from the Southern Stellar Stream Spectroscopic Survey ($S^5$) as well as proper motion measurements from $Gaia$'s third data release (DR3), and apply a Bayesian mixture modelling approach to isolate high-probability member stars. We recover extra-tidal features similar to those found in \citet{ShippEtal2018} surrounding each cluster. We conduct N-body simulations and compare the expected spatial distribution and variation in the dynamical parameters along the orbit with those of our potential member sample. Furthermore, we use Dark Energy Camera (DECam) photometry to inspect the distribution of the member stars in the color-magnitude diagram (CMD). We find that potential members agree reasonably with the N-body simulations and that the majority of them follow a simple stellar population-like distribution in the CMD which is characteristic of GCs. We link the extra-tidal features with their orbital properties and find that the presence of the tails agrees well with the theory of stellar stream formation through tidal disruption. In the case of NGC~1904, we clearly detect the tidal debris escaping the inner and outer Lagrange points which are expected to be prominent when at or close to the apocenter of its orbit.
Our analysis allows for further exploration of other GCs in the Milky Way that exhibit similar extra-tidal features.
\nopagebreak
} 


\maketitle
%
\section{Introduction}
\label{sec:introduction}

Globular clusters (GCs) are densely packed, spheroidal distributions of stars bound together by their gravity. Milky Way GCs can mostly be found within its halo or the bulge in orbit around the Galaxy. During a GC's orbit, the Galactic tidal interaction leads to a gradual loss of member stars. According to \citet{Reina-Campos2020} who used 25 present-day Milky Way-mass zoom simulations from the E-MOSAICS project to estimate the fraction of globular cluster stars within the halo, $\sim 0.3 \%$ of the stars in the halo can be attributed to disrupted globular clusters. Meanwhile, $2.3 \%$ can be attributed to other star clusters, and the majority of the halo is rather formed by the accretion of dwarf galaxies. Observationally, around 2\% of halo field stars show the chemical abundance anomalies of second-generation globular cluster stars \citep{MartellGrebel2010, MartellEtal2011, MartellEtal2016, SchiavonEtal2017, KochEtal2019, HortaEtal2021}. Converting this into a fraction of the halo that originally formed in globular clusters is strongly influenced by the model one uses for the star formation history in a globular cluster and the extent of early first-generation mass loss, but estimates have ranged from 11\% \citep{HortaEtal2021} to 50\% \citep{MartellGrebel2010}.

The gradual loss of mass that GCs undergo can be the result of several mechanisms. In the outskirts of a GC, stars can be stripped away by the Galaxy's potential therefore gaining an energy above the critical energy needed for their escape from the GC's potential. This leads to the formation of a diffuse stellar envelope surrounding the cluster beyond its tidal radius \citep{FukushigeHeggie2000, Baumgardt2001, ClaydonEtal2017, DanielEtal2017}.  Other mechanisms are possible that lead to the ejection of stars from within GCs \citep{LeighSills011, GrondinEtal2023}. These processes include natal kicks \citep{MerrittEtal2004}, ejections via 
supernova events \citep{ShenEtal2018, KounkelEtal2022}, or three-body encounters that can result in the ejection of one of the stars leaving behind a binary \citep{StoneLeigh2019, ManwadkarEtal2020, MontanariEtal2022, GrondinEtal2023}. The main depopulation mechanism of GCs is the escape of stars through the Lagrange points, formed by
the combined potentials of a given GC and the Galaxy \citep[e.g.][]{Kuepper+2008,WeatherfordEtal2023, XuEtal2024}. This mechanism results in the formation of long streams of stars sometimes referred to as tidal tails \citep[e.g.,][and references therein]{IbataEtal2021}.

Uncovering these streams is of great importance as they provide a means to probe the acceleration field of the Galaxy \citep[e.g.][]{Johnston+1999,Ibata+2001, FellhauerEtal2009} which can then constrain the dark matter distribution within the halo. Given the hierarchical merging scenario, the detection of stellar streams also reveals past merger events. Additionally, dark matter subhalos can be studied using stellar streams by examining gaps in the streams that could have resulted from impacts with these halos \citep[e.g.][]{IbataEtal2002, JohnstonEtal2002, Carlberg2012,Bovy2016,Erkal+2016, BanikEtal2018, BonacaEtal2019b, MontanariEtal2022} allowing for tests of $\Lambda$CDM itself.

Tidal tails of GCs have complex morphologies, and the orientation of the tails is correlated with the phase of orbit around the Galaxy -- especially for GCs with eccentric orbits \citep{CapuzzoDolcettaEtal2005, MontuoriEtal2007}. Refer to Figure 5 of \citet{MontuoriEtal2007} for a schematic of all forces acting on a tidally disrupted GC that lead to the formation of tidal tails.
Palomar 5 is a well studied example of a GC with prominent tidal tails \citep{Odenkirchen2001, Odenkirchen2003, Ibata+2016,BonacaEtal2020}. The outer parts of the tails of Palomar 5 align with the direction of the orbit. The leading tail has a slightly lower energy and thus orbits closer to the Milky Way, while the trailing tail has a higher energy and thus orbits further from the Milky Way, both tails forming the characteristic S-shape centered on the cluster. Since the location of the Sun is above the orbital plane of Palomar 5, the difference in the tails' distances from the Galactic center is projected onto the sky and is thus easily detectable. 
The orientation of the tails of GCs is also correlated with the location of the GC along its orbit and its eccentricity. The outer tails, or parts of the tidal tails that are farther away from the cluster, always align with the path of the orbit. On the other hand, the inner tails, i.e. parts of the tidal tails close to the GC, roughly align with the orbital path only when the cluster is near the pericenter \citep{CapuzzoDolcettaEtal2005, MontuoriEtal2007}. When the cluster approaches apocenter, the inner tails point in the direction of the Galactic center and anti-center. This distinction between the orientation of the inner and outer tails when the cluster is at apocenter then gives the impression that the GC has multiple tidal tails surrounding it. The same morphology dependence has been demonstrated for dwarf galaxies accreted onto the Milky Way in \citet{KlimentowskiEtal2009}. Finally, a GC that has spent a long time in the Galactic halo will undergo several orbital periods, potentially leading to the production of many streams within the halo. This mechanism also leads to observing more than one pair of streams when probing an area on the sky surrounding such a GC. 
Some GCs that have been found with such structures surrounding them include NGC~288 \citep{LeonEtal2000, Sollima2020} and NGC~2298 \citep{BalbinotEtal2011}.

In this work we explore two other GCs,  NGC~1261 and NGC~1904, which have also shown multiple stream-like structures surrounding them. Both GCs are relatively metal-rich with mean metallicities of $-1.33 \pm 0.02$ and $-1.66 \pm 0.01$ respectively \citep{WanEtal2023} and are thought to have been accreted alongside the Gaia-Enceladus event \citep{LimbergEtal2022}. NGC~1261 has a present-day mass of $1.67\times 10^{5}$~M$_{\odot}$ with a heliocentric distance of 16.4~kpc and Galactocentric distance of 18.28~kpc \citep{Baumgardt+2018}.  Based on the simulations of \citet{WanEtal2023}, the orbital period of NGC 1261 is $\sim 170$~Myr, with a pericenter distance of $\sim 2$~kpc, and an apocenter distance of $\sim 20$~kpc. With such an orbital period, NGC~1261 is likely to have completed more than $10$ full orbits around the galaxy within $2$~Gyrs and lost a large portion ($\sim50\%$) of its mass due to stellar evolution and the frequent interaction with the Galaxy's tidal field \citep[e.g.][]{BalbinotGiels2017}. NGC~1904 (M~79), has a present-day mass of $1.69\times 10^{5}$~M$_{\odot}$ with heliocentric and Galactocentric distances of $13.08$~kpc and $19.1$~kpc respectively \citep{Baumgardt+2018}. It has a small pericenter with a Galactocentric radius of $\sim 0.33$~kpc where the tidal effect from the Galaxy is strong. At the present, this GC lies very close to the apocenter of its orbit.

\citet{LeonEtal2000} pointed out the existence of a tidal tail for NGC~1261 oriented in the direction of the Galactic center, and works including \citet{KuzmaEtal2017} and \citet{WanEtal2023} pointed out the existence of a stellar envelope surrounding this cluster. Similarly, an extended halo of stars was found around NGC~1904, first in \citet{GrillmairEtal1995} and also in \citet{LeonEtal2000}, \citet{ZhangEtal2022}, \citet{WanEtal2023} and \citet{XuEtal2024}.  In \citet{ShippEtal2018}, data from the first three years of the Dark Energy Survey \citep[DES:][]{AbottEtal2018} were used to identify stellar streams in the Milky Way halo. 
Their results also included evidence for extra-tidal features around four Milky Way GCs, namely  NGC~288, NGC~1261, NGC~1851, and NGC~1904. \citet{ShippEtal2018} found stellar overdensities associated with NGC~1261 and NGC~1904 that do not align with the general direction of their orbits and in fact form a cross-shaped pattern centered on the clusters. Though they have pointed out these overdensities, they left further exploration to future work.
More recently, \citet{Sollima2020} has found three extra-tidal structures surrounding NGC~1904 that point in multiple directions, and \citet{IbataEtal2023} suggested that two streams could be associated to NGC~1261.  In this paper, we focus on these two GCs, and perform the subsequent analysis hinted at by \citet{ShippEtal2018} to provide further clarity pertaining to the origin of the clusters' extra-tidal features. 

Given the detection of these features in the aforementioned works, and the current location of NGC~1904 along its orbit, it seems possible that multiple stream-like features can be associated to these GCs, though more work is required to make that connection. Consequently, the Southern Stellar Stream Spectroscopic Survey \citep[$S^5$]{LiEtal2019}, has performed follow-up observations of these two clusters and in the fields surrounding them to expand on the findings of \citet{ShippEtal2018}. The survey has targeted regions on the sky along the expected orbit of the clusters as well as in the regions where the extra-tidal features from \citet{ShippEtal2018} were found. The survey provides radial velocities as well as metallicity measurements to help constrain extra-tidal member stars associated with these clusters in the targeted fields. 

The goal of this paper is to explore the cross-shaped features reported by \citet{ShippEtal2018} around the GCs NGC~1261 and NGC~1904, making use of the deep spectroscopic observations of $S^5$. 
We employ a Bayesian mixture modelling technique that incorporates proper motion measurements as well as radial velocity and metallicity values to isolate the GC and stream stars from the surrounding field stars within an area spanning $\sim10r_J$ where $r_J$ is the Jacobi radius of each cluster. After extracting the studied structures, we compare their spatial distribution and properties to what is expected when running N-body simulations of the two GCs. We primarily attempt to link the substructure found surrounding the clusters to their orbits around the Galaxy.

This paper is organized as follows. In Section~\ref{sec:Data}, we explain the data collected by $S^5$, specifically the target selection during the follow-up observations and the processing that the measurements undergo. In Section~\ref{sec:methods}, we describe the methodology applied to the selected sample of stars for each GC to distinguish high probability member stars. We also explain how the N-body simulations are constructed for comparison with our results. Section~\ref{sec:results} describes our results pertaining to the detected streams or substructure detected around the studied GCs. We then discuss these results in Section~\ref{sec:discussion}. In particular, we discuss the reliability of our methodology as well as the formation mechanisms of any uncovered substructure. Section~\ref{sec:conclusion} then summarizes our results and presents future prospects of this work. 

\begin{figure}
  \centering
  \includegraphics[width=\columnwidth]{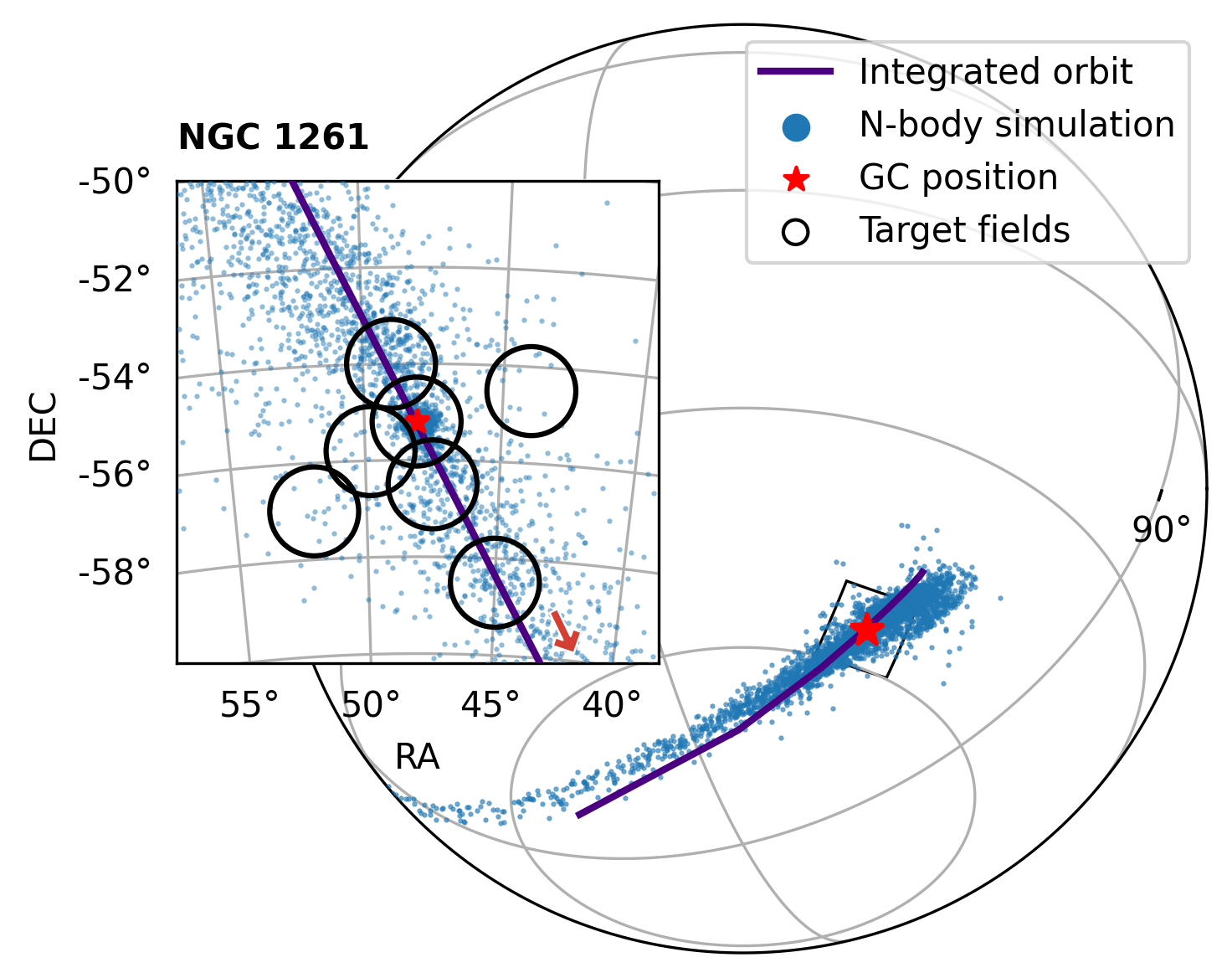} \includegraphics[width=\columnwidth]{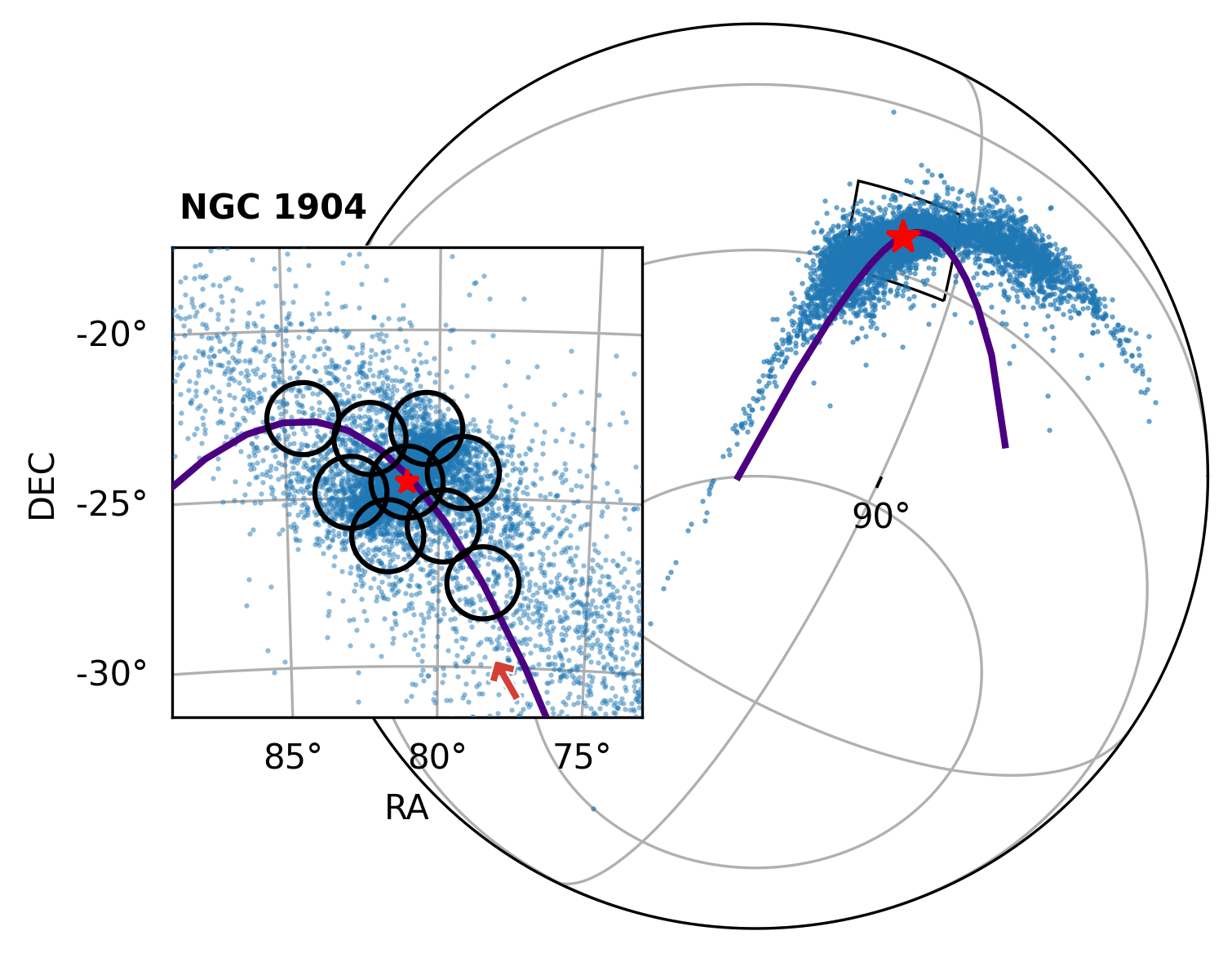} 
  \caption{3D sky plots for NGC~1261 (top) and NGC~1904 (bottom). The position of each GC is shown as a red star and the particles from the N-body simulations are shown in blue throughout the paper. 
  The zoomed panels show the distribution of particles close to the clusters and the outlines of the fields targeted by $S^5$ (black circles). We also plot in purple the integrated orbit of each GC given the gravitational potential described in Section~\ref{sec:simData} and the direction of motion of the cluster is indicated by the red arrow.}
  \label{fig:Skyplots}
\end{figure}

\section{Data}
\label{sec:Data}



Observations of these two GCs were taken as part of $S^5$, which was initiated in 2018 using the Two-degree Field (2dF) fiber positioner \citep{Lewis:2002} coupled with the dual-arm AAOmega spectrograph \citep{Sharp:2006} on the 3.9-m Anglo-Australian Telescope (AAT). This ongoing survey pursues a complete census of known streams in the Southern hemisphere. GCs and dwarf galaxies experiencing tidal disruption are also targeted, e.g.\ the Crater II and Antlia II dwarf galaxies \citep{Ji2021}.
We refer the reader to \citet{LiEtal2019} and \citet{LiEtal2022} for a detailed description of the instrument setup, observations, data reduction, and validation for $S^5$. The first public data release (DR1) was made available in 2021 \citep{S5DR1}, containing data collected from 2018 to 2020. The specific catalog used in this analysis is based on an internal data release, iDR3.7 of $S^5$, which is slated to become the second public data release (DR2) in 2025, with a more detailed description to be provided therein (T.S.~Li et al., in prep).

We provide a brief summary of the reduction pipeline here. As described in $S^5$ DR1 \citep{LiEtal2019, S5DR1}, the {\ttfamily rvspecfit} pipeline \citep{Koposov2019} is used to determine radial velocity (RV) and other stellar parameters, such as metallicity ([Fe/H]), effective temperature, and surface gravity for each star. The pipeline uses the PHOENIX-2.0 high-resolution stellar spectra library~\citep{Husser2013}, truncates the template spectra to the AAOmega wavelength range, and convolves them to match the AAOmega spectral resolution ($R\sim1,300$ for the blue arm at 3700--5700~\AA\ and $R\sim10,000$ for the red arm at 8400--8800~\AA), followed by multidimensional interpolation between templates. The fitting process involves a Nelder-Mead optimization to find the maximum likelihood point in the space of stellar atmospheric parameters and RVs, with subsequent Markov Chain Monte Carlo sampling to derive the full posterior distribution for each parameter. Moreover, the data reduction pipeline for DR2 has undergone significant improvements compared to DR1, most notably through the simultaneous modeling of multiple spectra, fitting both the blue-arm and red-arm spectra, as well as repeated observations of the same object from different nights, with proper consideration of the heliocentric correction for each observation.


The observations for these two GCs were mostly taken in 2020-2021, with some added in 2023. The GC targets are selected using parallax and proper motions from either \emph{Gaia} DR2 (for observations taken before Dec 2020) or \emph{Gaia} DR3. For parallax, we select stars with $\varpi - 3\sigma_{\varpi}<0.2$ to remove nearby disk contaminants. For proper motions, we select targets with $|\vec{\mu}-\vec{\mu}_0| < 1.6~{\rm mas~yr}^{-1}$, where $\vec{\mu}_0$ is the proper motion of each GC from \citet[][for observations taken before March 2021]{Vasiliev2019} and \citet[][for observations taken after March 2021]{Vasiliev_Baumgardt_2021}.
The targets were further selected to have similar color and magnitude as known GC members using the photometry from DES DR2 \citep{Abbott2021ApJS}. Specifically, dereddened $g$ and $r$ magnitudes are used for target selection, using the dust maps from \citet{SchlegelEtal1998}. 
A total of 7 AAT fields were observed for NGC 1261, and 9 fields for NGC 1904 (Figure~\ref{fig:Skyplots}). Observations of the central regions \citep{WanEtal2023} were added to the $S^5$ observations and processed using the same data reduction pipeline. 



The radial velocities are converted from heliocentric to the galactic standard of rest (gsr) using the latest galactocentric frame parameters from {\ttfamily astropy v5.1}. We also use the coordinate system $(\phi_1, \phi_2)$ defined using the rotation matrices provided in Appendix~\ref{appendix} for the two GCs, where $\phi_1$ is directed along the expected orbit of the clusters and $\phi_2$ is orthogonal to that direction. We select stars with radial velocity $-250<v_{gsr}/{\rm km~s}^{-1} <250$
and metallicity [Fe/H]$<0$ to remove stars with radial velocities and metallicities very far away from any known measurements for the GCs. Additionally, we enforce quality cuts to keep stars that have good measurements by the survey, i.e.\ stars with the flag {\ttfamily good\_star}~$=1$, and stars with small measurement errors on the radial velocity and metallicity (i.e.\ {\ttfamily vel\_calib\_std}~$<20$ ${\rm km~s}^{-1}$ and {\ttfamily feh\_calib\_std}~$<1$). In total, we retain 938 and 1706 stars in the 7 and 9 AAT fields surrounding NGC~1261 and NGC~1904 respectively. We subsequently apply the procedure described in Section~\ref{sec:methods} to analyse these samples.

\section{Methods}
\label{sec:methods}

\subsection{N-body simulations}
\label{sec:simData}

In order to compare the observations with theoretical predictions, we perform $N$-body simulations of both GCs. This approach is similar to \cite{Massari+2024}, who studied NGC~6254 and NGC~6397 with \textit{Euclid}. These simulations are performed with the $N$-body part of \textsc{Gadget-3}, which is an improved version of \textsc{Gadget-2} \citep{Springel_2005}. We model the Milky Way gravitational potential using the three component \texttt{MWPotential2014} model from \cite{Bovy_2015}. This consists of an NFW halo \citep{Navarro+1996} with a mass of $8\times10^{11}$~M$_\odot$, a scale radius of $16$ kpc, and a concentration of 15.3, a Miyamoto-Nagai disk \citep{Miyamoto_Nagai1975} with a mass of $6.8\times10^{10}$~M$_\odot$, a scale radius of $3$~kpc, and a scale height of $0.28$~kpc, and a broken power-law bulge with a mass of $5\times10^9$~M$_\odot$, a power-law exponent of $-1.8$, and a cutoff radius of $1.9$~kpc. 

In addition to the Milky Way, we also include the Large Magellanic Cloud (LMC), which has been shown to have a significant effect on the orbits and physical properties of streams in the Milky Way \citep[e.g.][]{Erkal+2019,Erkal+2020,Patel+2020,Vasiliev+2021,Pace+2022,Correa-Magnus+2022, KoposovEtal2023}. We take a similar approach to \cite{Erkal+2019} and model the LMC as a particle sourcing a Hernquist potential \citep{Hernquist1990} with a mass of $1.5\times10^{11}$~M$_\odot$ and a scale radius of $17.13$~kpc motivated by the LMC mass measured in \cite{Erkal+2019}. We include the effect of dynamical friction on the LMC using the results of \cite{Jethwa+2016}. We also account for the reflex motion of the Milky Way in response to the LMC. This is done using the approach of \cite{Vasiliev+2021} where the simulation is performed in the non-inertial Milky Way frame and the acceleration of the LMC on the Milky Way is subtracted off the LMC and the GC particles. As a sanity check, we integrated the GC orbits using the approach of \cite{Erkal+2019} where the Milky Way is included as a particle that moves in response to the LMC and we verified that the orbits relative to the Milky Way are identical.

In order to perform our simulations, we first rewind NGC~1261 and NGC~1904 in the combined potential of the Milky Way and the LMC for 2 Gyr. The present-day phase space coordinates of each GC (i.e.\ on-sky angles, proper motions, radial velocity, and distance) are taken from \citet{Vasiliev_Baumgardt_2021}. For the LMC's present-day phase-space coordinates we use the proper motions, distance, and radial velocity from \cite{Kallivayalil+2013,Pietrzynski+2019} and \citet{van_der_Marel+2002} respectively. After rewinding, we track the orbit of each GC and determine when they reach apocenter. For both GCs, we choose to simulate them forward from 4 apocenters ago. For NGC 1904 and NGC 1261 this corresponds to 0.97 Gyr and 1.19 Gyr ago respectively. We choose to inject the GCs at apocenter to give them time to adjust to the tidal field before pericenter.

We then inject the GCs at their respective locations and simulate them forward to the present day in the combined potential of the Milky Way and the LMC. Each GC is modeled as a King profile with $10^5$ particles where the profile can be characterized by the profile normalization factor $W$, the tidal radius $r_t$, and the mass $m$. We extract from \cite{de_Boer+2019} the values of $W$ and $r_t$ and from \cite{Baumgardt+2018} the values for the masses of each cluster. For NGC 1261, we use $W=6.8$, $r_t=37.81$~pc, and a mass of $1.67\times10^{5}$~M$_\odot$. For NGC 1904, we use $W=7.93$, a tidal radius of $r_t=38.32$~pc, and a mass of $1.69\times10^{5}$~M$_\odot$. For the softening, we simulate both GCs in isolation for 2 Gyr with a range of softening parameters, $(2.5,5,10,20)\times10^{-2}$~pc, and find that a softening of $5\times10^{-2}$~pc maintains the initial King profile with the smallest change. Both GCs are simulated forward to the present day, and both experience significant tides during their pericentric passages with the Milky Way\footnote{Videos for the NGC 1904 and NGC 1261 simulations are available at \url{https://www.youtube.com/watch?v=R-zooau62hk} and \url{https://www.youtube.com/watch?v=Gr8legg0QeQ} respectively.}. Both clusters end up close to their present-day location, with offsets of only $0.18^{\circ}$ and $0.57^{\circ}$ for NGC 1904 and NGC 1261 respectively. We therefore correct for this offset by shifting the position of the stars in right ascension $\alpha$ and declination $\delta$ so that the GCs match their observed present-day positions. This slight shift has a minimal effect on the proper motions and radial velocities of the simulated particles. We note that in these simulations, NGC 1904 and NGC 1261 lose 11\% and 2\% of their mass due to tidal stripping. Since this mass loss is comparable to the uncertainty on their present mass \citep[][]{Baumgardt+2018}, we do not iterate the procedure to match their present-day mass.

In Figure~\ref{fig:Skyplots}, we provide 3D sky plots for NGC~1261 (top) and NGC~1904 (bottom). The red star indicates the current position of the GCs, and the blue dots indicate the results of the N-body simulations. We also overplot the outline of the measured fields done by $S^5$ for both of these clusters in the zoom plot shown in the top and bottom panels. Furthermore, we show the expected orbit of each globular cluster integrated using the same gravitational potential described above. The direction of the orbit is indicated by the arrow. The simulations will be used to compare the properties of the stars classified as probable cluster members and for further discussion in Section~\ref{sec:discussion}.

\subsection{Mixture model description}


We aim to detect members of each GC separately, within the regions observed by $S^5$ which extend beyond the tidal radius of each cluster as explained in Section~\ref{sec:Data}. We therefore employ a mixture modelling approach whereby we disentangle stars that have similar properties to the clusters from the surrounding field stars (hereafter referred to as the background). Such probabilistic approaches have been used previously such as in the work of \citet{Sollima2020} for searching for tidal tails around a sample of 18 globular clusters, or \citet{KuzmaEtal2021} for finding tidal extensions to Omega Centauri. Similarly, the works of \citet{PaceLi2019} and \citet{McConnachieVenn2020} apply a probabilistic modelling approach on data from \emph{Gaia} DR2 to study ultra-faint dwarf galaxies. Earlier work includes \citet{WalkerEtal2009}, \citet{KoposovEtal2011}, and \citet{WalkerPenarrubia2011} who use a probabilistic approach for studying the Milky Way's dwarf spheroidal galaxies. Similarly to \citet{KuzmaEtal2021}, in this work we model the different properties of the GCs with Gaussian mixture modelling while fitting for the parameters that optimize a log-likelihood function within a Bayesian framework.

\begin{table*}[t]
\caption{Prior range and best-fit values for the mixture model parameters for both NGC~1261 and NGC~1904.}
\centering          
\begin{tabularx}{\textwidth}{@{\extracolsep{\fill}}
crcccccc@{}
} 

\toprule
\thead{Parameters}
&\thead{Priors}
&\thead{Ranges}
&\thead{Best-fit values}
&\thead{Errors}
&\thead{Ranges}
&\thead{Best-fit values}
&\thead{Errors} \\

& & NGC~1261 & NGC~1261 & NGC~1261 &  NGC~1904 & NGC~1904 & NGC~1904\\

\toprule
\textbf{Stream component} \\


$f_{\rm in}$ & Uniform & $(0,1)$ & 0.705 & 0.036 & $(0,1)$ & 0.639 & 0.030\\
$f_{\rm out}$ & Uniform & $(0,1)$ & 0.046 & 0.010 & $(0,1)$ & 0.034 & 0.005\\
$v_1$ & Uniform & $(-160,-80)$ & -84.037 & 0.381 & $(35, 41)$ & 38.283 & 0.333\\
$v_2$ & Uniform & $(0, 125)$ & 72.216 & 8.855 & $(-40, 125)$ & 1.355 & 6.412 \\
$v_3$ & Uniform & $(-300, 100)$ & -78.271 & 23.547 & $(-300, 100)$ & -54.988 & 17.449\\
$\sigma_{v, s}$ & Log-Uniform & $(0.1, 10)$ & 2.822 & 0.316 & $(0.1, 10)$ & 3.057 & 0.301 \\
$\mu_{\alpha,1}$ & Uniform & $(1, 2)$ & 1.599 & 0.007 & $(2, 3)$ & 2.471 & 0.008\\
$\mu_{\alpha,2}$ & Uniform & $(-1, 1)$ & -0.393 & 0.187 & $(-1, 1)$ & 0.236 & 0.148\\
$\mu_{\alpha,3}$ & Uniform & $(-3, 3)$ & 1.564 & 0.574 & $(-1.5, 4)$ & 0.835 & 0.412\\
$\sigma_{\mu_{\alpha}, s}$ & Log-Uniform & $(0.01,10)$ & 0.025 & 0.010 & $(0.01, 10)$ & 0.067 & 0.008\\
$\mu_{\delta,1}$ & Uniform & $(-2.15, 0)$ & -2.066 & 0.011 & $(-3, 0)$ & -1.591 & 0.009\\ 
$\mu_{\delta,2}$ & Uniform & $(-1, 3)$ & 1.679 & 0.292 & $(-1, 2)$ & -0.160 & 0.180 \\
$\mu_{\delta,3}$ & Uniform & $(-5, 5)$ & -1.877 & 0.807 & $(-5, 2)$ & -0.671 & 0.510\\
$\sigma_{\mu_{\delta}, s}$ & Log-Uniform & $(0.01,10)$ & 0.053 & 0.014 & $(0.01, 10)$ & 0.064 & 0.008\\
$\overline{[\mathrm{Fe/H}]}_s$ & Uniform  & $(-4,0)$ & -1.285 & 0.023 & $(-4, 0)$ & -1.605 & 0.021\\
$\sigma_{\mathrm{[Fe/H]}, s}$ & Log-Uniform & $(0.001, 1)$ & 0.196 & 0.020  & $(0.001, 1)$ & 0.210 & 0.017\\
\midrule
\textbf{Background component} \\

$v_{bg}$ & Uniform & $(-100, 50)$ & -48.175 & 3.049 & $(-60, 50)$ & -39.708 & 2.540\\
$\sigma_{v, bg}$ & Log-Uniform & $(0.1, 1000)$ & 83.650 & 2.108 & $(0.1, 1000)$ & 98.725 & 1.812\\ 
$\mu_{\alpha, bg}$ & Uniform & $(0, 3)$ & 1.860 & 0.024 & $(0, 3)$ & 2.330 & 0.018\\ 
$\sigma_{\mu_{\alpha}, bg}$ & Log-Uniform & $(0.1, 100)$ & 0.667 & 0.018 & $(0.1, 100)$ & 0.684  &  0.013\\
$\mu_{\delta, bg}$ & Uniform & $(-4, 0)$ & -1.840 & 0.028 & $(-4, 0)$ & -1.529 & 0.018\\
$\sigma_{\mu_{\delta}, bg}$ & Log-Uniform & $(0.01,100)$ & 0.768 & 0.021 & $(0.01,100)$ & 0.701 & 0.013\\
$\overline{[\mathrm{Fe/H}]}_{bg}$ & Uniform & $(-5, 1)$ & -1.315 & 0.020 & $(-5, 1)$ & -1.002 & 0.014\\
$\sigma_{\mathrm{[Fe/H]}, bg}$ & Log-Uniform & $(0.1, 100)$ & 0.480 & 0.015 & $(0.1, 100)$ & 0.493  & 0.011 \\
\bottomrule
\end{tabularx}
\tablefoot{Radial velocity and its dispersion is in units of ${\rm km~s}^{-1}$, and proper motions and their dispersions are in units of ${\rm mas~yr}^{-1}$. A horizontal line has been added to separate between parameters fitting for the stream component and those fitting for the background component. The dispersions mentioned all correspond to the intrinsic dispersion of the respective parameter distribution and not the total one.}
\label{tab:table1}
\end{table*}

We assume that for each star we have the following information:
star positional coordinates  $\phi_1$ and $\phi_2$, proper motions along right ascension and declination ($\mu_{\alpha}$ and $\mu_{\delta}$ respectively), radial velocity $v$ and the $[\mathrm{Fe/H}]$ abundance.  Note that $\mu_{\alpha}$ always contains the $\cos{\delta}$ term. We also assume  knowledge of the measurement errors 
$\sigma_{\mu_{\alpha},meas}$,
$\sigma_{\mu_{\delta},meas}$,
$\sigma_{v,meas}$
and
$\sigma_{[\mathrm{Fe/H}],meas}$ for the above quantities, as well as the correlation coefficient $\rho$ linking the measurement noise of the two proper motions.
We will build a density model
for $u = (\mu_{\alpha}, \mu_{\delta}, v, [\mathrm{Fe/H}])$, given the positional information $\phi_1$ and measurement noise characteristics collected in
$\zeta = (\sigma_{\mu_{\alpha},meas},
\sigma_{\mu_{\delta},meas},
\sigma_{v,meas},
\sigma_{[\mathrm{Fe/H]},meas},\rho)$. The use of $\phi_2$ will be defined later in the modelling.

Our two-component mixture model $p(u\ |\ \phi_1, \zeta)$ 
mixes a stream component $p_s(u\ |\ \phi_1,\zeta)$ with a background component $p_{bg}(u\ |\ \zeta)$ in the following way:
\begin{equation}
p(u\ |\ \phi_1,\zeta) = f \cdot p_s(u\ |\ \phi_1,\zeta) + (1-f) \cdot p_{bg}(u\ |\ \zeta) \enspace ,
\label{eq:mixture}
\end{equation}
where $0\le f \le 1$ is the mixture coefficient of the stream component.
The stream component models properties of stars belonging to GCs, while the background component models other contaminant stars within the field. In both GC and background component models the distribution over $u$ is factorized into three independent factors - two univariate ones over radial velocity and metallicity, $p_v$ and $p_{\mathrm{[Fe/H]}}$, respectively, and a bivariate one, $p_{pm}$, for proper motions.
All component factors will be formulated as Gaussians fully specified by their means and (co)variance structures.
 
%
%
The kinematics along the stream vary as a function of the position on the sky, and this variation can be small or large depending on how far or close a GC is from its turning points on the orbit. To include this variation in our modeling, the mean $v_s$ of the velocity factor of the stream mixture component $p_{v,s}$ will therefore be modelled as 
a quadratic function $v_s(x) = v_1 + v_2 x + v_3 x^2$ of the position within the stream, $x = \phi_1 /10^{\circ}$, parametrized by the coefficients $(v_1,v_2,v_3)$. 
A similar dependence on the stream-position can be argued for the mean 
$\vec{\mu}_s(x) = (\mu_{\alpha, s}(x) ,\mu_{\delta, s}(x) )^T$
of the proper motion factor $p_{pm, s}$ of the stream component:
\begin{eqnarray}
\mu_{\alpha, s}(x) 
&=&  \mu_{\alpha, 1}  + \mu_{\alpha, 2} x + \mu_{\alpha, 3} x^2 \enspace , \\
\mu_{\delta, s}(x) 
&=&  \mu_{\delta, 1} + \mu_{\delta, 2} x +  \mu_{\delta, 3} x^2 \enspace ,
\end{eqnarray}
parameterized by the coefficients
$(\mu_{\alpha, 1}, \mu_{\alpha, 2}, \mu_{\alpha, 3},\mu_{\delta, 1}, \mu_{\delta, 2}, \mu_{\delta, 3})$.

No such dependence on the stream-position needs to be captured by the background component, hence, the means $v_{bg}$ and 
$\vec{\mu}_{bg} =\left(\mu_{\alpha, bg}, \mu_{\delta, bg}\right)$ of the velocity and proper motion factors, respectively, of the background mixture component can be considered constant parameters of our mixture model.

Finally, the evolution of a GC within the Galaxy does not change its initial metallicity. Hence, the mean metallicity for both the stream,  and background components, $\overline{[\mathrm{Fe/H}]}_s$ and $\overline{[\mathrm{Fe/H}]}_{bg}$, respectively, are considered constant parameters.

Having specified the mean structure of our mixture model, we now focus on the (co)variance one. For the univariate ($v$ and [Fe/H]) and bivariate factors ($\mu_{\alpha}$ and $\mu_{\delta}$), the variance is a superposition of the intrinsic and measurement dispersions.
Hence,
\begin{equation}
\sigma_j = \sqrt{\sigma_{j,int}^2 + \sigma^2_{j,meas}} \enspace ,
\end{equation}
where $j$ is an element of $u$. Again, the intrinsic dispersions are free parameters of the model. For proper motions, we also impose a full bivariate Gaussian model with covariance structure
\begin{equation}
\centering
\begin{matrix}{\Sigma} = 
\begin{pmatrix}
 \sigma_{\mu_{\alpha}}^2 & \rho \sigma_{\mu_{\alpha}, meas}\sigma_{\mu_{\delta}, meas} \\
\rho \sigma_{\mu_{\alpha}, meas}\sigma_{\mu_{\delta}, meas} & \sigma_{\mu_{\delta}}^2
\end{pmatrix} \enspace.    
\end{matrix}
\end{equation}

The above treatments are done for both the stream and background component which will be indicated by the subscripts $s$ and $bg$ where needed. We are now ready to fully specify the model of eq. \eqref{eq:mixture}:
\begin{equation}
p_s(u\ |\ \phi_1,\zeta) = 
p_{v,s}(v\ |\ \phi_1,\zeta) \cdot
p_{\mathrm{[Fe/H]}, s}([\mathrm{Fe/H}]\ |\ \zeta) \cdot
p_{pm, s}(\vec{\mu}\ |\ \phi_1,\zeta) \; ,   
\end{equation}
where each of the factors are provided here:
\begin{equation}
p_{v,s}(v\ |\ \phi_1,\zeta) =  \frac{1}{\sigma_{v,s} \sqrt{2 \pi}} \exp{\left[-\frac{1}{2} \Bigg(\frac{v - v_s(x)}{\sigma_{v,s}}\Bigg)^2\right]} \enspace ,   
\end{equation}
\begin{multline}
p_{\mathrm{[Fe/H]}, s}([\mathrm{Fe/H}]\ |\ \zeta) =  \\
    \frac{1}{\sigma_{\mathrm{[Fe/H]}, s} \sqrt{2 \pi}} \exp{\left[-\frac{1}{2} \Bigg(\frac{[\mathrm{Fe/H}] - \overline{[\mathrm{Fe/H}]}_s}{\sigma_{\mathrm{[Fe/H]}, s}}\Bigg)^2\right]} \enspace ,
\end{multline}
\begin{multline}
p_{pm, s}(\vec{\mu}\ |\ \phi_1,\zeta) = \\
\frac{1}{\sqrt{(2 \pi)^2 \ |\ \Sigma_s\ |\ }} \exp{\left[-\frac{1}{2} \Big(\vec{\mu}- \vec{\mu}_s(x)\Big)^{T} \Sigma_s^{-1} \Big(\vec{\mu}- \vec{\mu}_s(x)\Big) \right]} \enspace .    
\end{multline}
For the background component we have a similar scheme except our parameters don't varry with $\phi_1$ but are constant values:
\begin{equation}
    p_{bg}(u\ |\ \zeta) = 
p_{v, bg}(v\ |\ \zeta) \cdot
p_{\mathrm{[Fe/H]}, bg}([\mathrm{Fe/H}]\ |\ \zeta) \cdot
p_{pm, bg}(\vec{\mu}\ |\ \zeta) \; ,
\end{equation}
where similarly the factors are detailed here:
\begin{equation}
    p_{v, bg}(v\ |\ \zeta) =  \frac{1}{\sigma_{v, bg} \sqrt{2 \pi}} \exp{\left[-\frac{1}{2} \Bigg(\frac{v - v_{bg}}{\sigma_{v, bg}}\Bigg)^2\right]} \enspace ,
\end{equation}

\begin{multline}
p_{\mathrm{[Fe/H]}, bg}([\mathrm{Fe/H}]\ |\ \zeta) = \\
    \frac{1}{\sigma_{\mathrm{[Fe/H]}, bg} \sqrt{2 \pi}} \exp{\left[-\frac{1}{2} \Bigg(\frac{[\mathrm{Fe/H}] - \overline{[\mathrm{Fe/H}]}_{bg}}{\sigma_{\mathrm{[Fe/H]}, bg}}\Bigg)^2\right]} \enspace ,    
\end{multline}

\begin{multline}
p_{pm, bg}(\vec{\mu}\ |\ \zeta) =  \\
\frac{1}{\sqrt{(2 \pi)^2 \ |\ \Sigma_{bg}\ |\ }} \exp{\left[-\frac{1}{2} \Big(\vec{\mu}- \vec{\mu}_{bg}\Big)^{T} \Sigma_{bg}^{-1} \Big(\vec{\mu}- \vec{\mu}_{bg}\Big) \right]} \enspace .
\end{multline}

There is a final nuance to our probabilistic modeling. We, in fact, have two models, one operating within the present-day Jacobi radius $r_J$ of the GC, the other outside of it, both identical up to a mixing coefficient or fraction $f$.  The separation is necessary as maintaining a single fraction of member stars for the whole setup would artificially increase the membership likelihood of stars within the stream component far away from the GC center. It is thus more realistic to assume two different fractions of member stars for the areas within and outside the Jacobi radius of each cluster. The two models can therefore be given by the following:
\begin{equation}
p_{\rm in}(u\ |\ \phi_1,\zeta) = f_{\rm in} \cdot p_s(u\ |\ \phi_1,\zeta) + (1-f_{\rm in}) \cdot p_{bg}(u\ |\ \zeta),
\label{eq:mixture_in}
\end{equation}
\begin{equation}
p_{\rm out}(u\ |\ \phi_1,\zeta) = f_{\rm out} \cdot p_s(u\ |\ \phi_1,\zeta) + (1-f_{\rm out}) \cdot p_{bg}(u\ |\ \zeta).
\label{eq:mixture_out}
\end{equation}
The coefficients $0<f_{\rm in}, f_{\rm out}<1$ are free model parameters.

Given a star with distance $r = \sqrt{\phi_1^2 + \phi_2^2}$ to the cluster's center and measurements
$u = (\mu_{\alpha}, \mu_{\delta}, v, [\mathrm{Fe/H}])$, together with positional information $\phi_1$ and measurement noise characteristics 
$\zeta = (\sigma_{\mu_{\alpha},meas},
\sigma_{\mu_{\delta},meas},
\sigma_{v,meas},
\sigma_{[\mathrm{Fe/H]},meas},\rho)$,
the overall model likelihood reads:
\begin{equation}
p(u\ |\ r, \phi_1,\zeta) = 
\left[ p_{\rm in}(u\ |\ \phi_1,\zeta) \right]^{I(r<r_J)}
\cdot
\left[ p_{\rm out}(u\ |\ \phi_1,\zeta) \right]^{1-I(r<r_J)},   
\end{equation}
where $I(r<r_J)$ is an indicator function equal to 1 when $r<r_J$ and 0 otherwise. Note that this is the only step that requires the knowledge of $\phi_2$ in the mixture model.

In total, we have 24 parameters that we attempt to fit. These consist first of the coefficients for the quadratic functions parametrizing the change of radial velocity and proper motions of the stream component: $\{ v_1, v_2, v_3\}$,  
$\{\mu_{\alpha, 1}, \mu_{\alpha, 2},  \mu_{\alpha, 3} \}$, and $\{\mu_{\delta,1}, \mu_{\delta,2}, \mu_{\delta,3} \}$, and the respective means for the background component: $\{v_{bg}, \mu_{{\alpha}, bg}, \mu_{\delta, bg}\}$. With regards to the metallicity, the mean stream and background metallicities $\overline{[\mathrm{Fe/H}]}_s$ and $\overline{[\mathrm{Fe/H}]}_{bg}$ respectively are fit. We also fit the intrinsic dispersions of the radial velocities, proper motions and metallicities for both the stream and background components: $\{ \sigma_{v,s}, \sigma_{\mu_{\alpha}, s}, \sigma_{\mu_{\delta}, s}, \sigma_{\mathrm{[Fe/H]}, s} \}$ and $\{ \sigma_{v, bg}, \sigma_{\mu_{\alpha}, bg}, \sigma_{\mu_{\delta}, bg}, \sigma_{\mathrm{[Fe/H]}, bg} \}$ respectively. We have dropped the subscript $int$ for better readability, but stress that we fit the intrinsic and not total dispersion. Finally, we also fit for the fraction of GC members parameter in and outside of the respective Jacobi radius of each GC, namely $f_{\rm in}$ and $f_{\rm out}$.

For the computation of the present day Jacobi radius, we use the result of \cite{King_1962}:

\begin{equation}
r_J = \left( \frac{Gm}{\Omega^2 - \frac{d^2\phi}{dR^2}}\right)^\frac{1}{3} ,
\end{equation}
where $m$ is the mass of the globular cluster, $\Omega$ is its angular velocity with respect to the Milky Way, and $\frac{d^2\phi}{dR^2}$ is the second derivative of the Milky Way's gravitational potential with respect to the distance from the Milky Way. We compute this Jacobi radius while rewinding each cluster's orbit in the presence of the Milky Way and the LMC. To account for uncertainties in the Milky Way and LMC potential, we use the same approach as \cite{Pace+2022} (see Section 3.1 of that work) who sampled over the Milky Way potential uncertainties using the results of \cite{McMillan2017}. We also sample over the uncertainties in the present-day phase-space of both clusters (i.e. their proper motions, distances, and radial velocities). For NGC 1261, we obtain for the present-day Jacobi radius $r_J = 170^{+4}_{-2}$~pc $\approx 0.594^\circ$ given a distance of $16.4$~kpc. For NGC 1904, we obtain $r_J = 175^{+3}_{-3}$~pc $\approx 0.764^\circ$ given a distance of $13.07$~kpc. These values are comparable to those detailed in \citet{BalbinotGiels2017}. Note that with this calculation, we can also extrapolate estimates of the pericenter of each cluster's orbit found to be $0.6 \pm 0.1$~kpc for NGC~1261 and $0.12 \pm 0.06$~kpc for NGC~1904. These values along with the other orbital parameters pertaining to these clusters will be discussed in Section~\ref{sec:discussion}.

\begin{figure*}[!ht]
  \centering
  \includegraphics[width=\columnwidth]{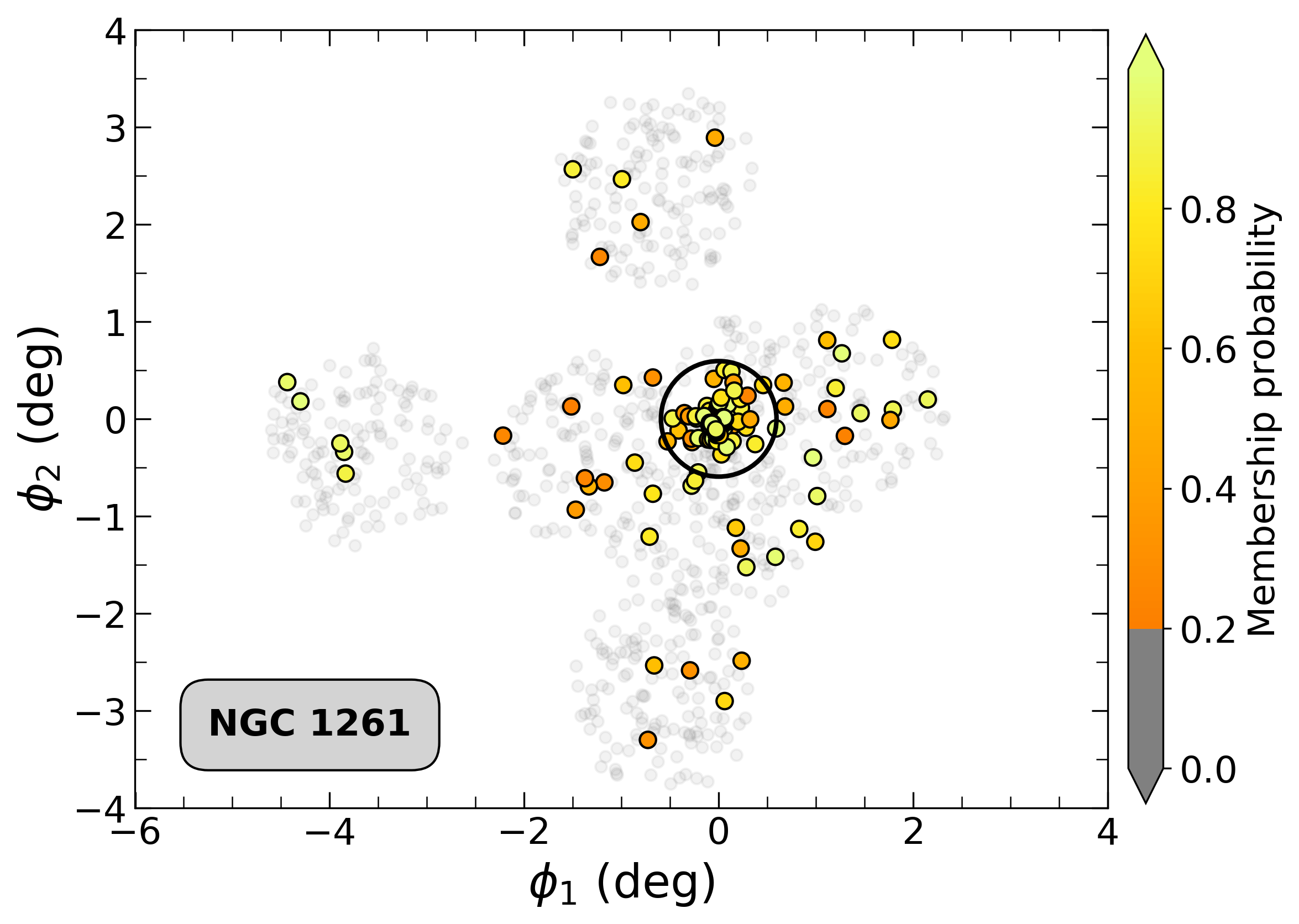} \includegraphics[width=\columnwidth]{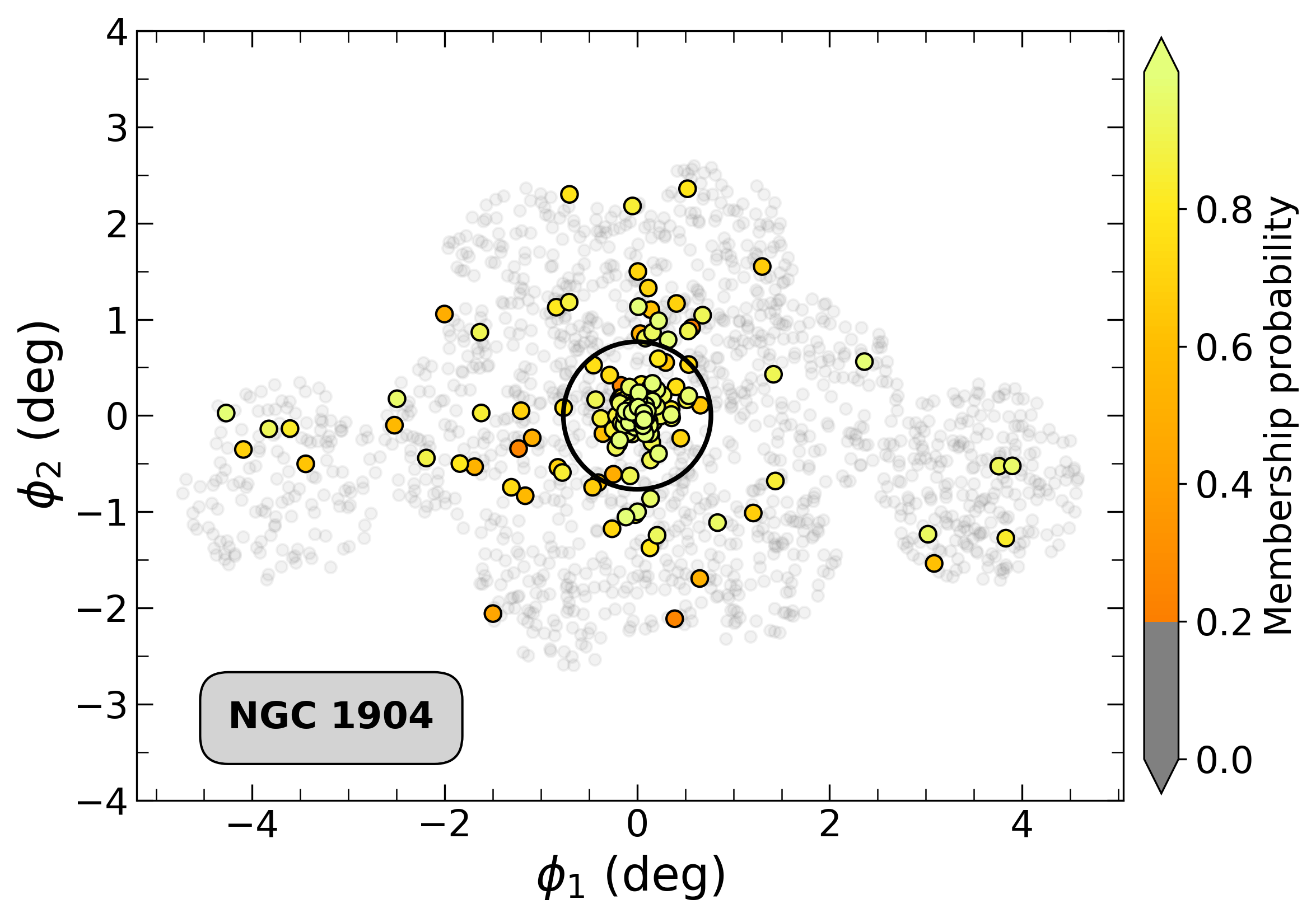} 
  \caption{Spatial distribution of the targeted stars, each colored by their membership probability of belonging to NGC~1261 (left) and NGC~1904 (right) respectively. All stars with probabilities less than 20\% are shown in grey. The solid circles outline the region within the Jacobi radius of each cluster and show that we detect many probable members outside of these regions.}
  \label{fig:probs_pos}
\end{figure*}

\begin{figure*}[!ht]
  \centering
\includegraphics[width=\columnwidth]{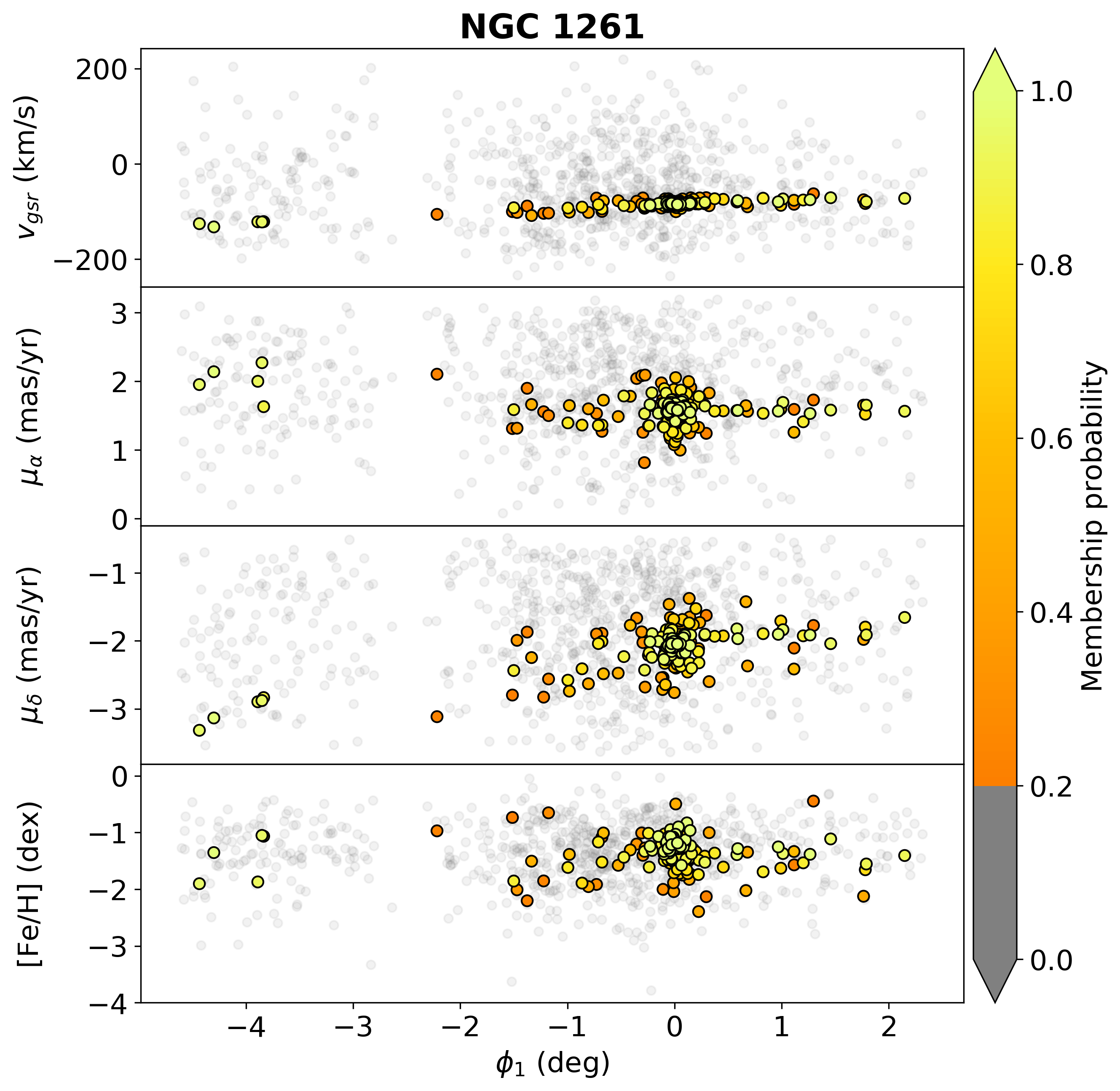} 
\includegraphics[width=\columnwidth]{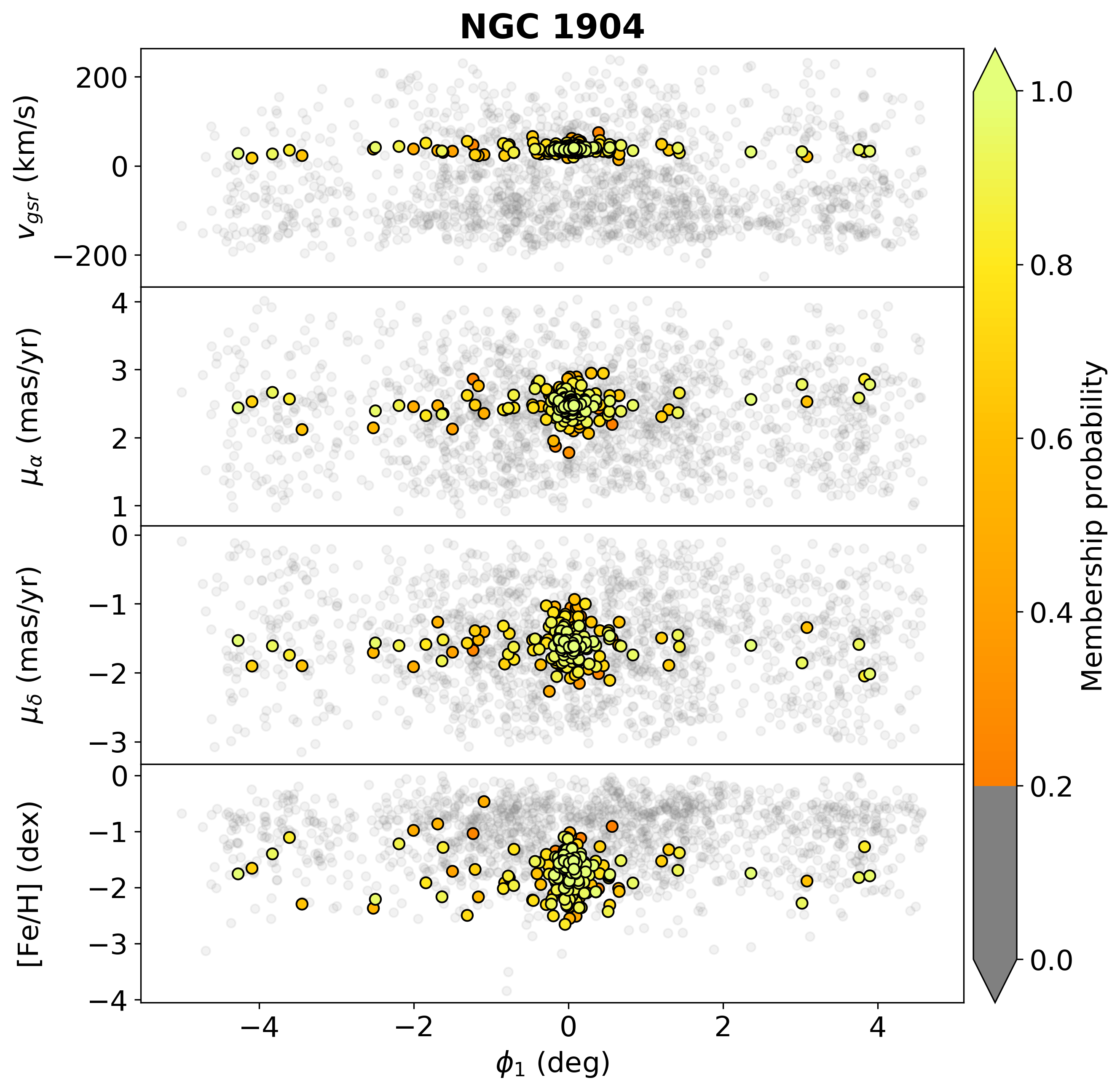} 
\includegraphics[width=0.95\columnwidth]{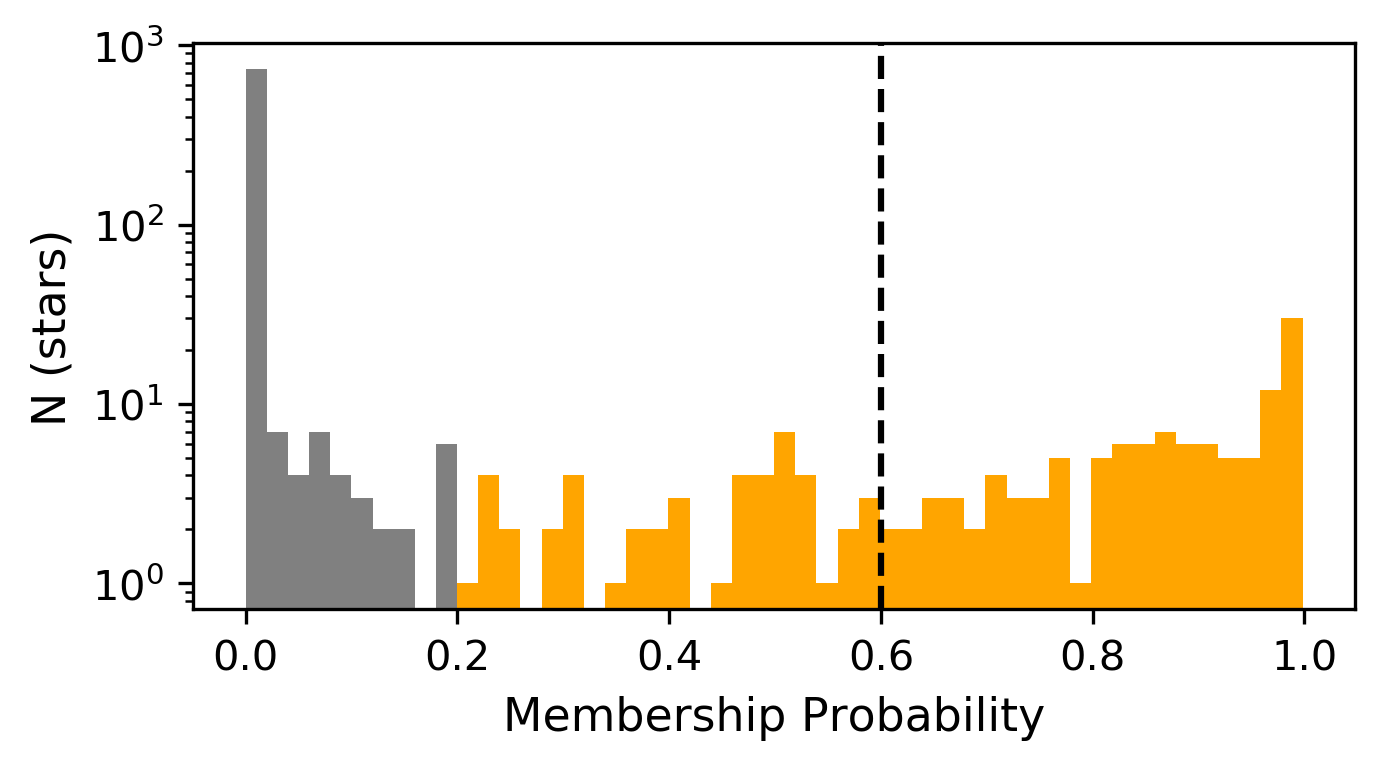} 
\hfill
\includegraphics[width=0.95\columnwidth]{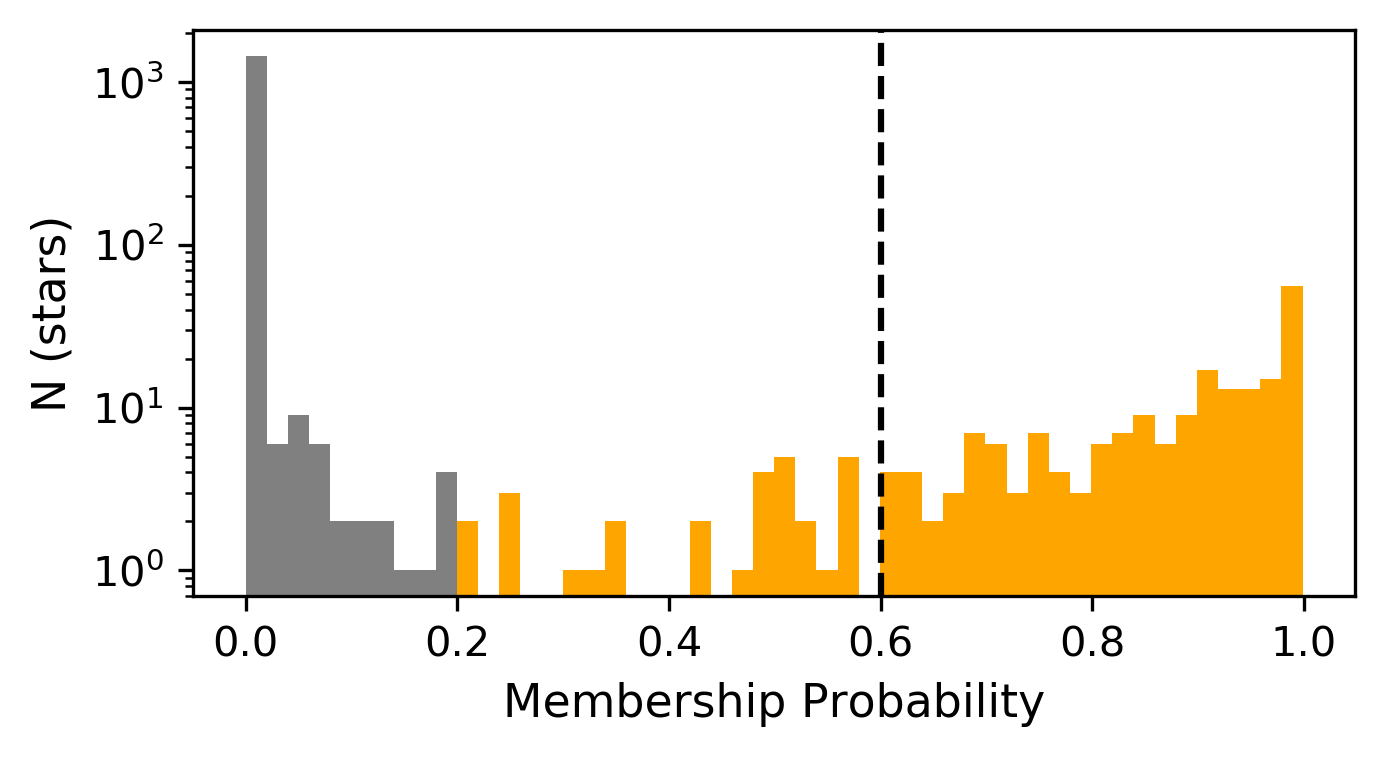} 
  \caption{Modelled properties of stars within the sample (left for NGC~1261 and right for NGC~1904) as a function of $\phi_1$. As in Figure~\ref{fig:probs_pos}, each point is colored by its membership probability of belonging to the respective cluster. The bottom two panels show the histograms of all probabilities in the sample. Vertical lines at 0.6 (60\%) represent the threshold on the probability.}
  \label{fig:probs}
\end{figure*}

\begin{figure*}
  \centering
\includegraphics[width=\textwidth]{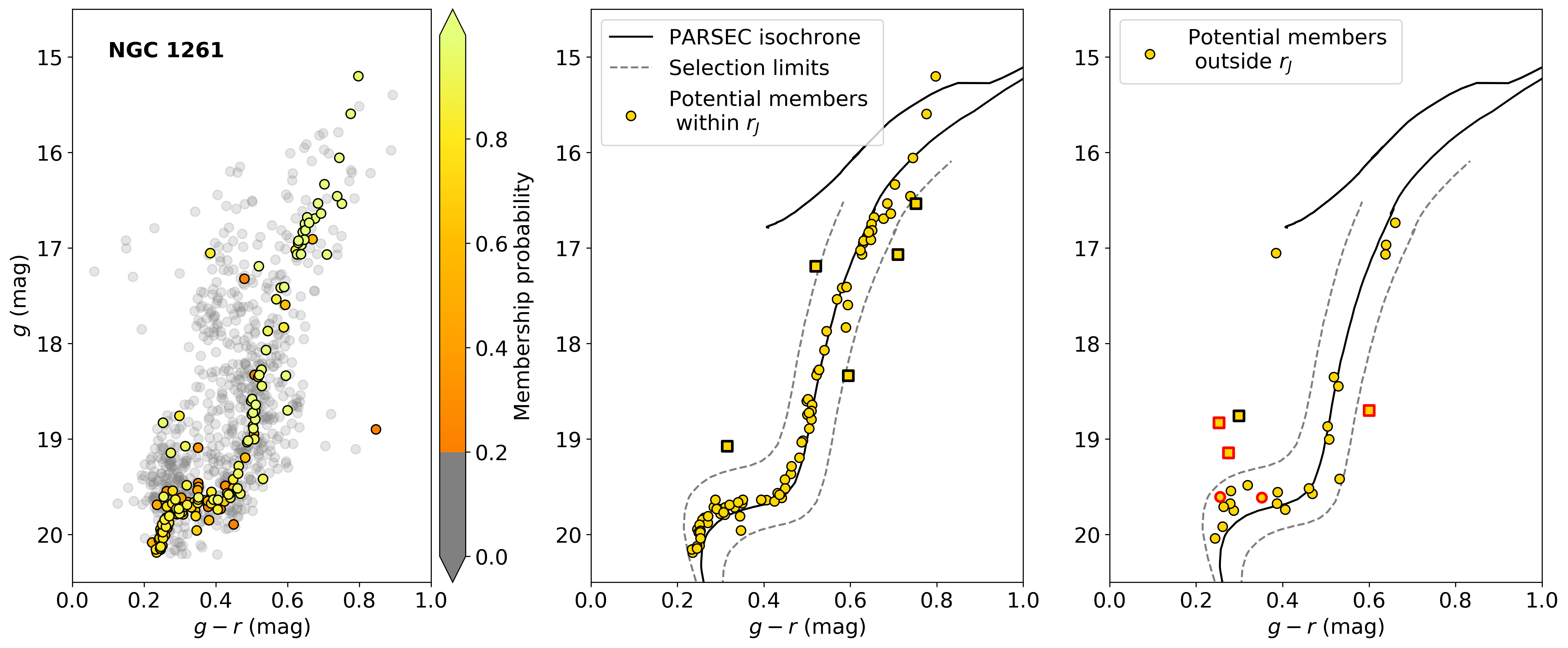} 
  \caption{The dereddened colour-magnitude diagram for likely NGC~1261 cluster and stream members using the $g$ and $r$ bands from the available DECam photometric measurements. Left panel: all stars in our sample for this cluster are color-coded by the calculated membership probability, and using grey for all stars with probabilities less than $20\%$. We observe a very narrow distribution that emerges within the whole sample. Middle panel: High-probability member stars are extracted using the threshold of $60\%$ and considering only stars within a radius $r=r_J$ around the cluster. An isochrone is also added for comparison. Dashed lines separate stars with similar CMD properties as the cluster from likely contaminants plotted using a square marker. Right panel: Similarly, we plot high-probability stars that lie outside of $r=r_J$. Stars outlined in red are the same stars selected in Figure~\ref{fig:results_1261} as having different $v_{gsr}$ from the N-body simulation prediction.}
  \label{fig:CMDs1261}
\end{figure*}

\begin{figure*}[ht!]
  \centering
\includegraphics[width=\textwidth]{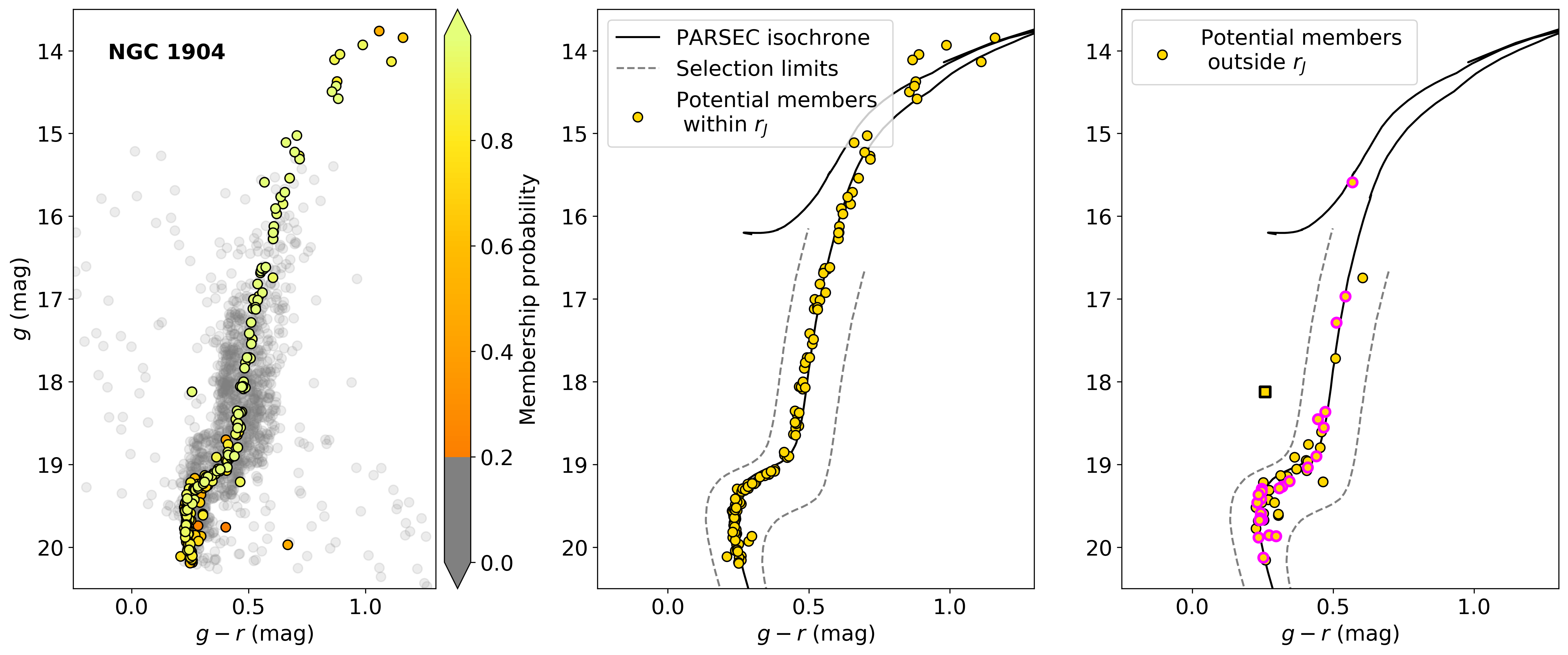} 
  \caption{As Figure~\ref{fig:CMDs1261} but for NGC~1904. The results for this GC show very little contamination as most potential members align well in the CMD. Stars outlined in magenta are the same as those selected in Figure~\ref{fig:results_1904} for their extension in $\phi_{2}$. These stars align well with the rest of the selection, indicating that they are likely to be true members of NGC~1904. }
  \label{fig:CMDs1904}
\end{figure*}

In fitting the mixture model parameters, we assume log-flat priors for the sampling of the dispersions and flat priors for all other parameters. In Table~\ref{tab:table1}, we provide the prior ranges assumed for each of the parameters in obtaining the results for NGC~1261 and NGC~1904. The sampling and optimization of the log-likelihood function is performed using the {\ttfamily emcee} package with 100 walkers and 3000 steps each for the burn-in process and the actual run. Once the best fit parameters are found, they are substituted within eqs. (1-19) and the GC membership probability for each star is calculated as the posterior probability of the star to belong to the stream component:

\begin{equation}\label{eq:MemberProb}
    P(s \ |\ r, u, \phi_1,\zeta)  = 
    \frac{f \cdot p_s(u\ |\ \phi_1,\zeta)}
{f \cdot p_s(u\ |\ \phi_1,\zeta) + (1-f) \cdot p_{bg}(u\ |\ \zeta)}
\end{equation}
where $f$ is either $f_{\rm in}$ or $f_{\rm out}$ depending on stars' positions with respect to $r_J$: 
\begin{equation}
f = \left(f_{\rm in}\right)^{I(r<r_J)} \cdot
\left(f_{\rm out}\right)^{1-I(r<r_J)}.    
\end{equation}

\begin{figure*}[ht!]
  \centering
\includegraphics[width=0.32\textwidth]{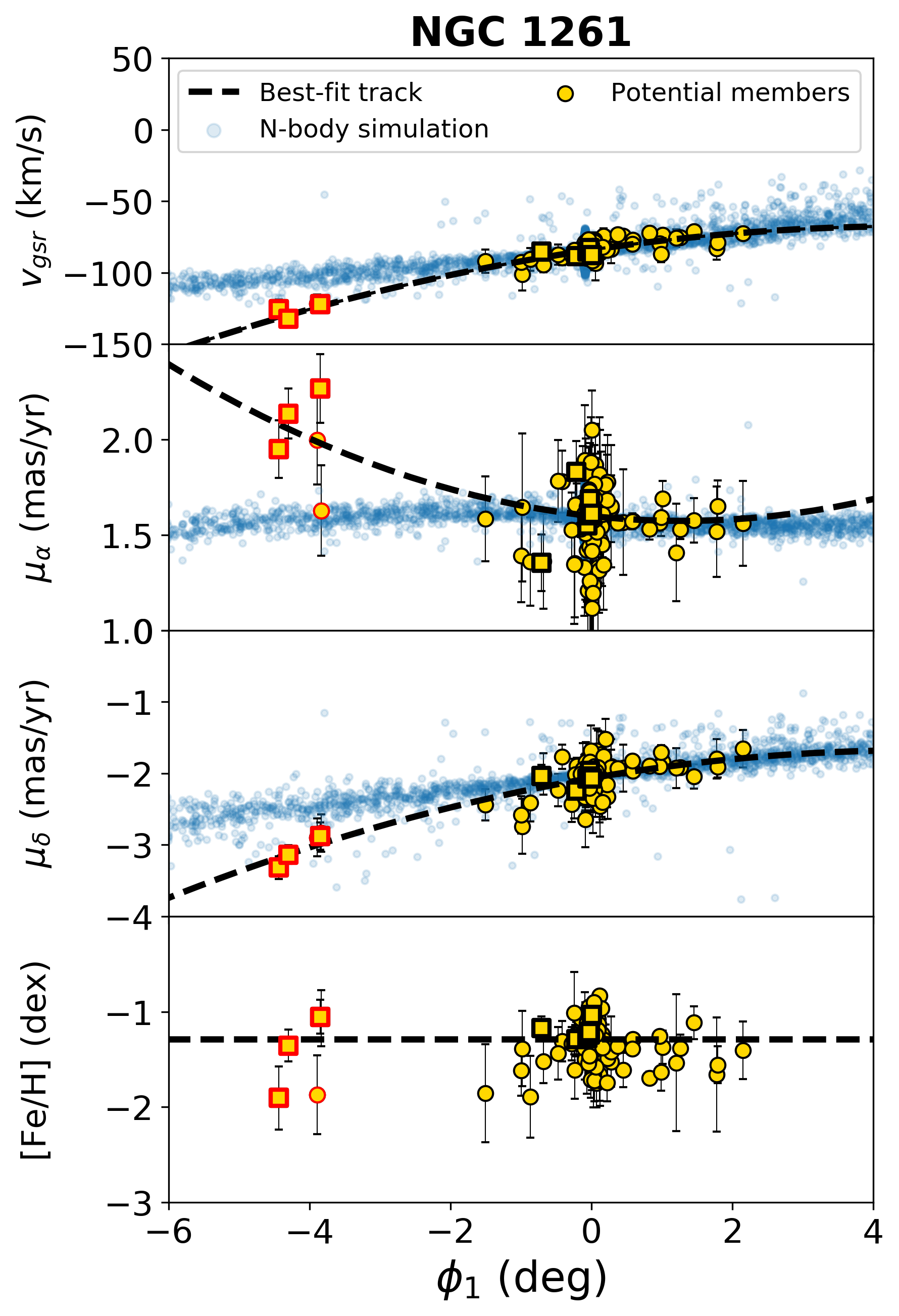} 
\includegraphics[width=0.67\textwidth]{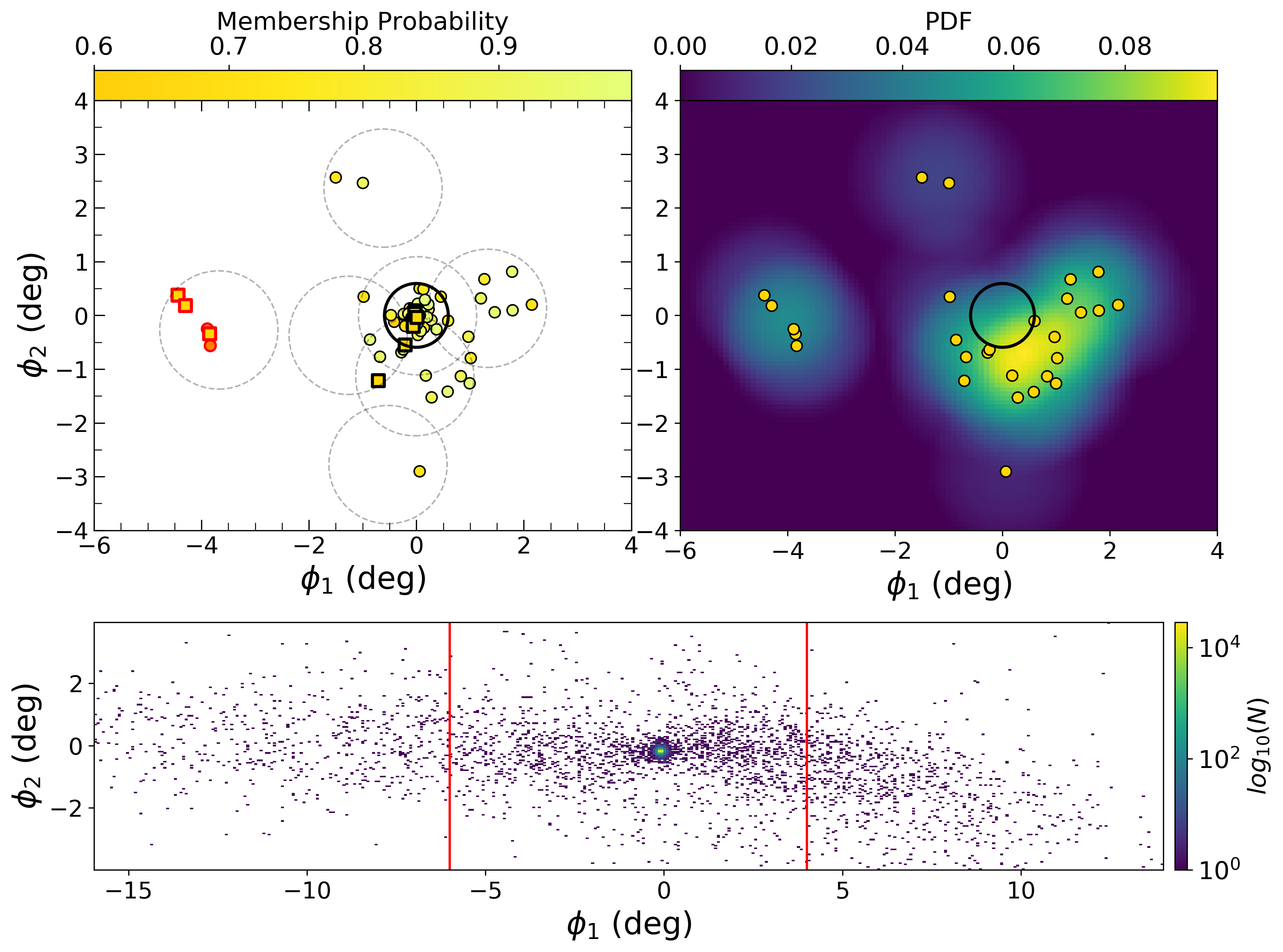} 
\caption{Left: Distribution in the modelled properties of the stars belonging to NGC~1261 and its stream as a function of $\phi_1$ after enforcing an $60\%$ lower limit on the membership probability. Error bars represent the measurement uncertainties. The black dashed line represents the best fit track used to model the mean value for the respective property at a given $\phi_1$. The results from the N-body simulation in each of the four dimensions are in blue. 
Stars with $v_{gsr}$ values inconsistent with the N-body simulation are outlined in red. Top middle: the spatial distribution of the same stars plotted in the left panel. Dashed circles show the outlines of the fields targeted by $S^5$. The solid circle shows the extent of $r_J$. Stars plotted using the square marker are those that showed different CMD properties from the cluster as shown in Figure~\ref{fig:CMDs1261}. Top right: 2D density plot of the distribution of potential member stars around NGC~1261. Stars within $r_J$ have been removed when creating this plot to enhance the density contrast.
Bottom right: 2D histogram showing the density of simulation particles on the sky. The vertical red lines indicate the same window on the sky as is shown in the top panels.}
\label{fig:results_1261}
\end{figure*}

By applying a threshold cut on this membership probability, we are able to separate stars that have a high likelihood of belonging to the GC from the background distribution of stars. In Table~\ref{tab:table1} we provide the best-fit values for the quantities we fit for, both for NGC~1261 and NGC~1904, as well as their respective errors. This includes our estimates for the velocities as well as the mean metallicities for each GC which are in excellent agreement with the literature. Further discussion around our estimates of the mean metallicity as well as the metallicity and velocity dispersions will be provided in Section~\ref{sec:properties}.

\section{Results}
\label{sec:results}

\begin{figure*}[ht!]
  \centering
\includegraphics[width=0.32\textwidth]{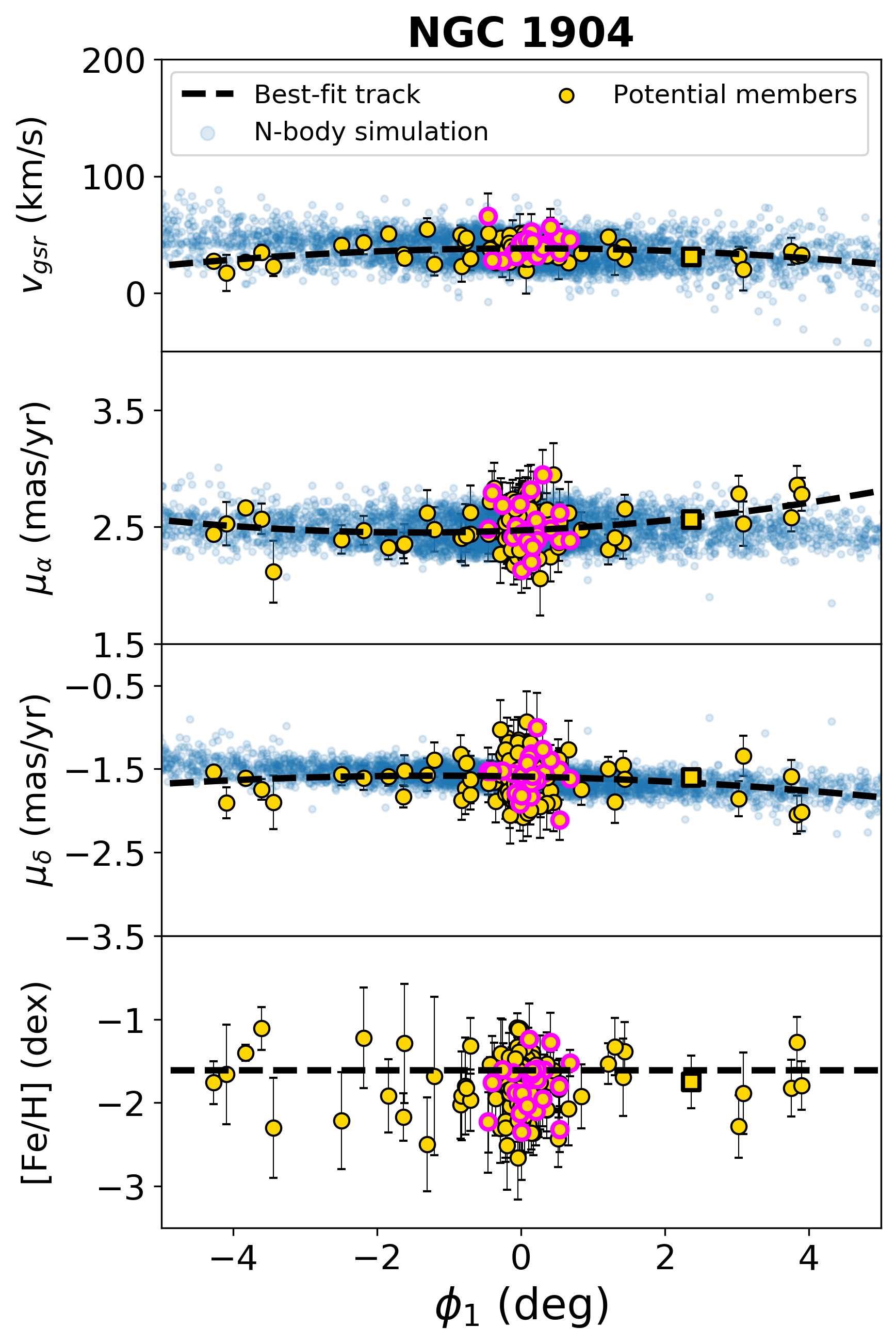} 
\includegraphics[width=0.67\textwidth]{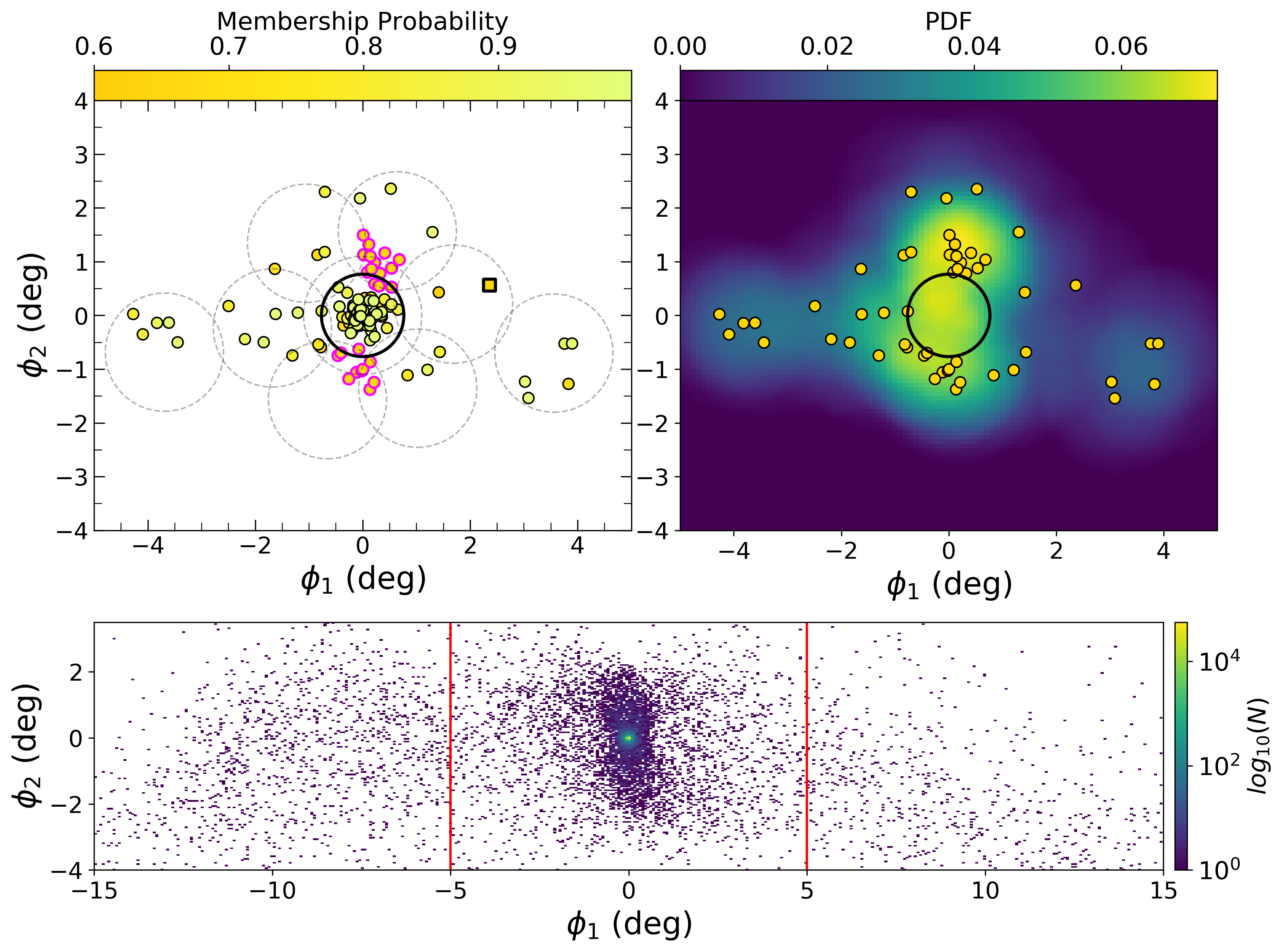} 
\caption{As Figure~\ref{fig:results_1261} but for NGC~1904. In the top middle panel, we select the stars that form overdensities above and below the cluster in $\phi_{2}$ in magenta. They are also highlighted in the left panel, showing that they have similar properties to the other likely cluster and stream stars. These overdensities are clearly visible in the density plot provided in the top right panel. The 2D histogram of the N-body simulation shown in bottom right panel also shows a clear distinction between the inner tails which correspond to our detected overdensities and the outer tails.}
\label{fig:results_1904}
\end{figure*}

In Figure~\ref{fig:probs_pos} we show the distribution of stars for each of our samples for NGC~1261 (left) and NGC~1904 (right). We color each star by the membership probability achieved after a full run of the methodology described in Section~\ref{sec:methods}. Stars with the highest probability of belonging to the respective GC are those in yellow. All stars with a membership probability less than $20\%$ are shown in solid grey. 

The first thing we observe is that the stars immediately surrounding the centers of the clusters are retrieved with very high probability ($>90\%$). 
Aside from the stars in the immediate vicinity of the GCs, we also detect high-membership probability stars within each of the fields observed by $S^5$ several degrees away from the clusters. 
For NGC~1261, we observe that the most probable member stars populate the target fields aligned with the direction of the GC's orbit around the Galaxy (i.e. direction of increasing $\phi_1$) as well as the fields that are orthogonal to the orientation of the orbit (direction of increasing $\phi_2$). 
This hints at the intersecting double stream feature that has been reported for this GC in works such as \citet{ShippEtal2018} and \citet{IbataEtal2023}. However, we caution that this could also be a result of limited sky coverage by $S^5$ such that the debris from NGC 1261 could have a much wider coverage and happens to be present in all the $S^5$ fields.

For NGC~1904, we also see probable member stars in all target fields observed by $S^5$. 
We observe the alignment of several high-probability stars extending both to the left and right of the GC, specifically stars with $\phi_1 < -1^{\circ}$ and $\phi_1 > 1^{\circ}$, indicating the presence of a stream of stars aligned with the direction of orbit of the GC. In addition to this feature, we also observe two groups of high-probability stars above and below the cluster in $\phi_2$ with $-1^{\circ}<\phi_1<1^{\circ}$. This is a sign 
of an inner stream to the GC referenced in \citet{GrillmairEtal1995}, \citet{Carballo-BelloEtal2018}, and \citet{ShippEtal2018}. We elaborate on these features for both GCs further later in this section and provide a discussion of their nature and origin in Section~\ref{sec:discussion}. 

In Figure~\ref{fig:probs} we display the distributions of $v_{gsr}$, $\mu_{\alpha}$, $\mu_{\delta}$ and $[\mathrm{Fe/H}]$ as a function of $\phi_1$ for all stars in our sample for NGC~1261 (left) and NGC~1904 (right). Each star is colored by the probability of belonging to the respective cluster as computed via eq. \eqref{eq:MemberProb} in Section~\ref{sec:methods}. 
In the bottom panels of the same figure, we also show the distributions of the membership probabilities for each GC. 
In order to isolate the stars that have a high probability of belonging to the respective GC, we enforce a threshold on the membership probability. By inspecting the probability distributions shown in the bottom panels of Figure~\ref{fig:probs}, we choose a threshold as high as possible as long as it doesn't eliminate clear members in the color-magnitude diagram (CMD, discussed further in Figures~\ref{fig:CMDs1261} and \ref{fig:CMDs1904}). 
This also corresponds approximately to the points at which we start seeing an increase in the number of stars with a larger probability in the bottom panels of Figure~\ref{fig:probs}. 
We therefore choose a threshold of $60\%$ as is indicated by the dashed line. Bins with probabilities $<20\%$ again are shown in grey. With this preliminary cut, we assess in what follows the true membership of the remaining stars by tracing their position in the CMD as well as comparing their spatial distributions and properties to what is expected from N-body simulations of the two GCs. We further discuss robustness of this technique and possible contamination in our samples in Section~\ref{sec:discussion}.

Another step is performed to ascertain the membership of the selected sample of stars to the two studied GCs, through inspecting their distribution within the CMD. To construct the CMDs of the two clusters, we make use of the bands available from DECam photometry, particularly the dereddened $g$ and $r$ bands which provide a much narrower CMD for the GC stars than what is possible using \emph{Gaia} photometry given the larger depth that DECam photometry is able to probe. In Figures~\ref{fig:CMDs1261} and \ref{fig:CMDs1904} we analyze the distributions in the CMD for NGC~1261 and NGC~1904 respectively. The left panel of these two figures shows all stars in our original sample colored by cluster membership probability. Similarly here, we plot all stars with probabilities less than $20\%$ in solid grey. We expect that for GCs, all member stars should follow a thin isochrone-like distribution in the CMD. It is therefore significantly noteworthy to see that the majority of the stars highlighted as high-probability members have such a distribution in the CMD, especially since no information about the colour or magnitude of the stars has been involved in the modelling. The middle and right panels show the remaining stars after enforcing the thresholds on the probability. In the middle panel, we show the potential members that lie within $r_J$ while the right panel shows all potential members outside of this region. We also plot PARSEC isochrones with mean metallicities taken from Table~\ref{tab:table1}, and ages 11.2 Gyrs for NGC~1261 and 11.7 Gyrs for NGC~1904 \citep{Baumgardt+2018}. These isochrones are also shifted using the respective distances of each GC i.e. 16.4~kpc and 13.08~kpc for NGC~1261 and NGC~1904 respectively. 

The middle panel shows that the majority of stars within the central region around the GC follow closely the isochrone. This is a good indication that the potential members we have in this region are true members. We also expect stream stars of a GC to follow the same isochrone. The right panel shows that many of the stars found in the stream indeed show this overlap with the isochrone though we see a slight scatter compared to the middle panel. The stars marked in red in Figure~\ref{fig:CMDs1261} will be further introduced and discussed through Figure~\ref{fig:results_1261}. Using the dashed line in the middle and right panels, we define regions where the stars outside $r_J$ have similar colour and magnitude properties as the respective clusters. These lines are defined such that the stars that are clearly far away from the isochrone are not included. Stars that lie outside the dashed lines are therefore contaminants and are indicated using a square marker. These stars are also plotted in the same way in Figures~\ref{fig:results_1261} to track their spatial positions and properties.

We present the same analysis for NGC~1904 in Figure~\ref{fig:CMDs1904}. Similarly, we observe a strikingly narrow isochrone-like distribution of the potential members in the CMD. Similar to NGC~1261, the potential members are divided between those that lie within and outside of $r_J$ and we observe near-perfect overlap between these stars and the GC isochrone. As for the stars outlined in magenta, they will be further introduced and discussed through Figure~\ref{fig:results_1904}. The stars outside of the chosen limits are shown by a square marker and are plotted in Figure~\ref{fig:results_1904} as well.

The remaining stars after enforcing our selection threshold are shown in Figures~\ref{fig:results_1261} and \ref{fig:results_1904} for NGC~1261 and NGC~1904 respectively. In both figures, we display the observed stars in yellow and what is expected from the N-body simulations in blue. The top middle panel of both figures shows the spatial distribution of the remaining stars. The dashed circles outline the target fields observed by $S^5$ and the solid circle shows the area outlined by the Jacobi radius. 
In the top right panel, we show the density plot for the remaining sample stars (shown again in yellow in this figure) by applying Kernel Density Estimation (KDE) using an Epaneshnikov kernel with $1.5^\circ$ for the bandwidth. Note that we take out the stars within $r_J$ of the GC in order to enhance the density contrast.

The bottom right panel in both figures shows a 2D histogram plot of what we would expect to see from our N-body simulations with the red vertical lines indicating the same area on the sky as the above panels. 
Lighter colors represent regions of higher density, while darker colors similarly represent regions of lower density. 
In the left panels, the dashed black lines show the best fit polynomial tracks defined in Section~\ref{sec:methods} that trace the mean variation of $v_{gsr}$, $\mu_{\alpha}$ and $\mu_{\delta}$ as a function of $\phi_1$. We emphasize that the polynomial tracks are solely a fit to the data without any assumption or knowledge of the GC orbits or Milky Way potential. In the bottom left panel, the horizontal dashed line similarly defines the best-fit mean metallicity $\overline{[\mathrm{Fe/H}]}_s$ as our estimate for the mean metallicity of each GC.

For NGC~1261 (Figure~\ref{fig:results_1261}), we find 88 potential members within $r_J$ and 28 outside $r_J$. We observe that the best fit track and the recovered member stars are quite similar to the results of the N-body simulation within $\sim 2^{\circ}$ of the cluster. The properties of the potential member stars also lie within the ranges defined by the simulation to within two standard deviations. The metallicity for each star is also consistent within two standard deviations from the best-fit mean metallicity of the cluster. For the region with $\phi_1 < -2.5^{\circ}$, however, the best-fit tracks begin to diverge from the expected distribution. Since the potential members in this region are questionable, we mark these stars with a red border in both the left and top right panels, and track them further in the rest of the analysis. With regards to the position of these stars in the CMD, we observe that two of the stars marked in red do not differ in their distribution within the CMD from the remaining potential members, showing that they have properties similar to those of the stars within $r_J$ from the cluster. The mismatch seen in Figure~\ref{fig:results_1261} between the expected properties of these stars in the simulations and their observed properties may imply that the Milky Way potential assumed for the N-body simulations is not accurate enough to reproduce the properties of the stream far away from the cluster. For example, we have ignored the effect of the Milky Way bar \citep[e.g.][]{Hattori+2016,Price-Whelan+2016,Erkal+2017,Pearson+2017}. Within $\pm 2^\circ$ from this GC however (roughly $4r_J$), we can confidently say that we detect highly probable member stars that are also expected from the simulations. We also discuss the divergence of the tracks in Section~\ref{sec:discussion}. In total, we observe a broad distribution of stars that overlaps with the regions where the cross-shaped structures were observed in \citet{ShippEtal2018}. Although this similarity could be a consequence of the fields chosen by $S^5$, we do confirm the existence of potential member stars in all fields. 

For NGC~1904 (Figure~\ref{fig:results_1904}), we find 143 potential members within $r_J$ and 51 outside $r_J$. We see that the best-fit tracks agree with the N-body simulation, and the likely cluster and stream members also have similar properties to what is expected from the simulation, considering the measurement uncertainties. We note the groups of stars on the top and bottom of the GC that seem to form an inner set of tidal tails that along with the horizontal distribution of stars along the cluster's orbit, give the appearance of "multiple-tidal tails" associated with this cluster. Such substructure is expected when a GC is very close to apocenter, which is the case for NGC~1904. To trace the properties of the potential member stars of the inner stream, we mark these stars with magenta outlines. These stars are also indicated in Figure~\ref{fig:CMDs1904} where we observe that they all align perfectly with the isochrone indicating that they are likely to be true members. The left panel of Figure~\ref{fig:results_1904} shows that the properties of these stars directly overlap with the properties of stars close to the GC. To further evidence the existence of the top and bottom overdensities of stars, they can be clearly seen in the top right panel as overdensities on the top and bottom of the cluster and a stream component along $\phi_1$. This distribution matches the results of \citet{ShippEtal2018} with respect to this cluster. This extension in the $\phi_{2}$ direction is also visible in the N-body simulations, as shown in the bottom right panel of Figure~\ref{fig:results_1904}. 
There too, the simulations predict the existence of an outer stream with $\phi_1 <-2^{\circ}$ or $\phi_1 >2^{\circ}$, and an inner stream in the region in between. We provide in Appendix~\ref{appendixB} some of the potential members identified in this work for NGC~1261 and NGC~1904 respectively with the full lists made available online. In the following section, we further discuss our methodology and the possible origin scenarios explaining the observed structures.

\section{Discussion}\label{sec:discussion}

\begin{figure*}
  \centering
\includegraphics[width=\columnwidth]{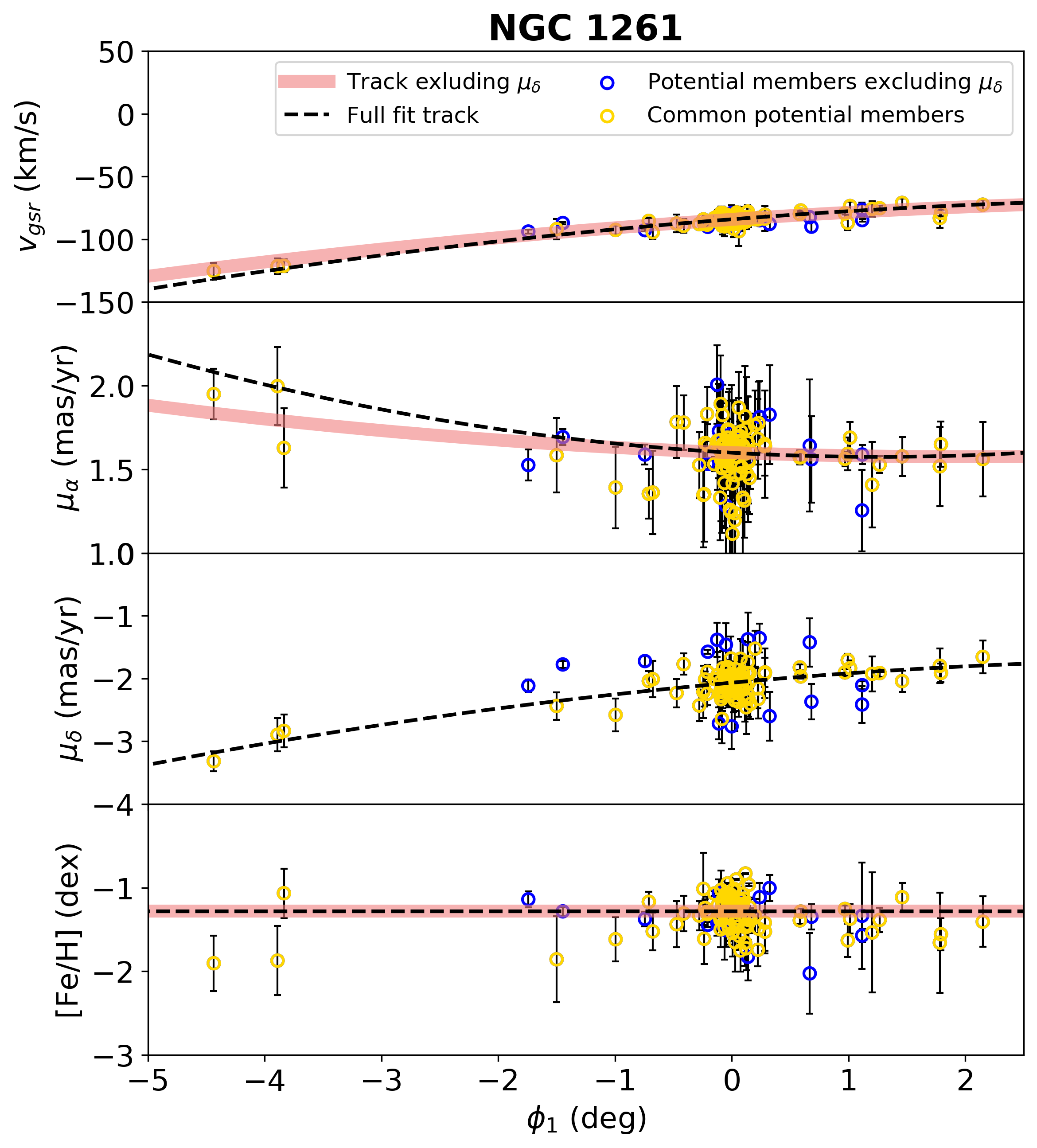} 
\includegraphics[width=\columnwidth]{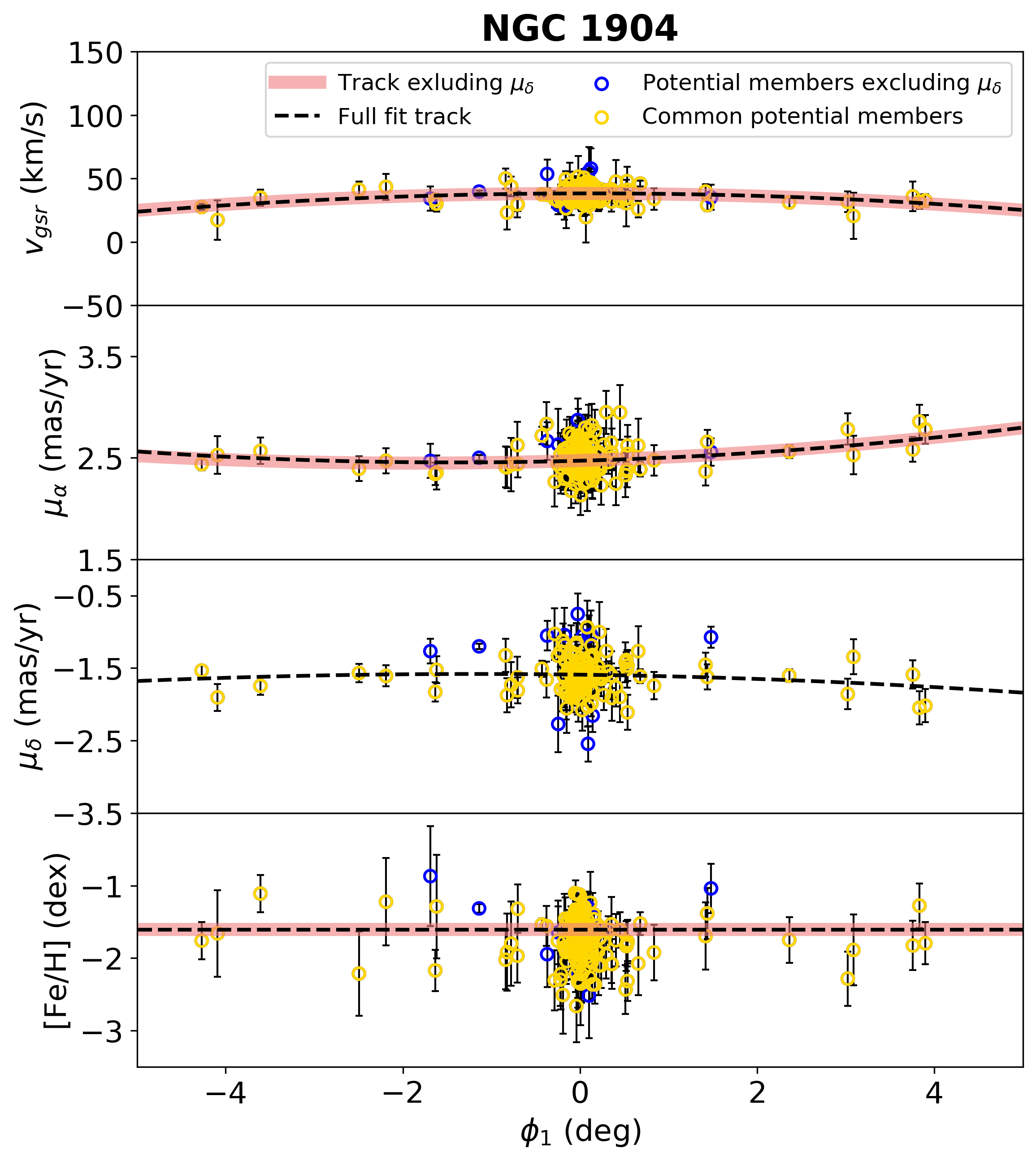} 
  \caption{A comparison between the full-fit runs described in Section~\ref{sec:methods}, and new runs where $\mu_{\delta}$ is not taken into account. The best fit tracks from the full-fit runs, first shown in Figures~\ref{fig:results_1261} and \ref{fig:results_1904}, are re-plotted here using the dashed line. The best-fit tracks from the new runs are shown using the solid red line. A new lower limit is enforced on the membership probability from the new runs and the remaining stars are cross-matched with the potential members from the full-fit results of Section~\ref{sec:results}. The common members between the two runs are shown in yellow and those that are not common are shown in dark blue. From this plot, we see that the distribution of stars in $\mu_{\delta}$ follows the same track as the original runs even though it was not taken into account.}
  \label{fig:overlap}
\end{figure*}

\begin{figure}[ht!]
  \centering
\includegraphics[width=\columnwidth]{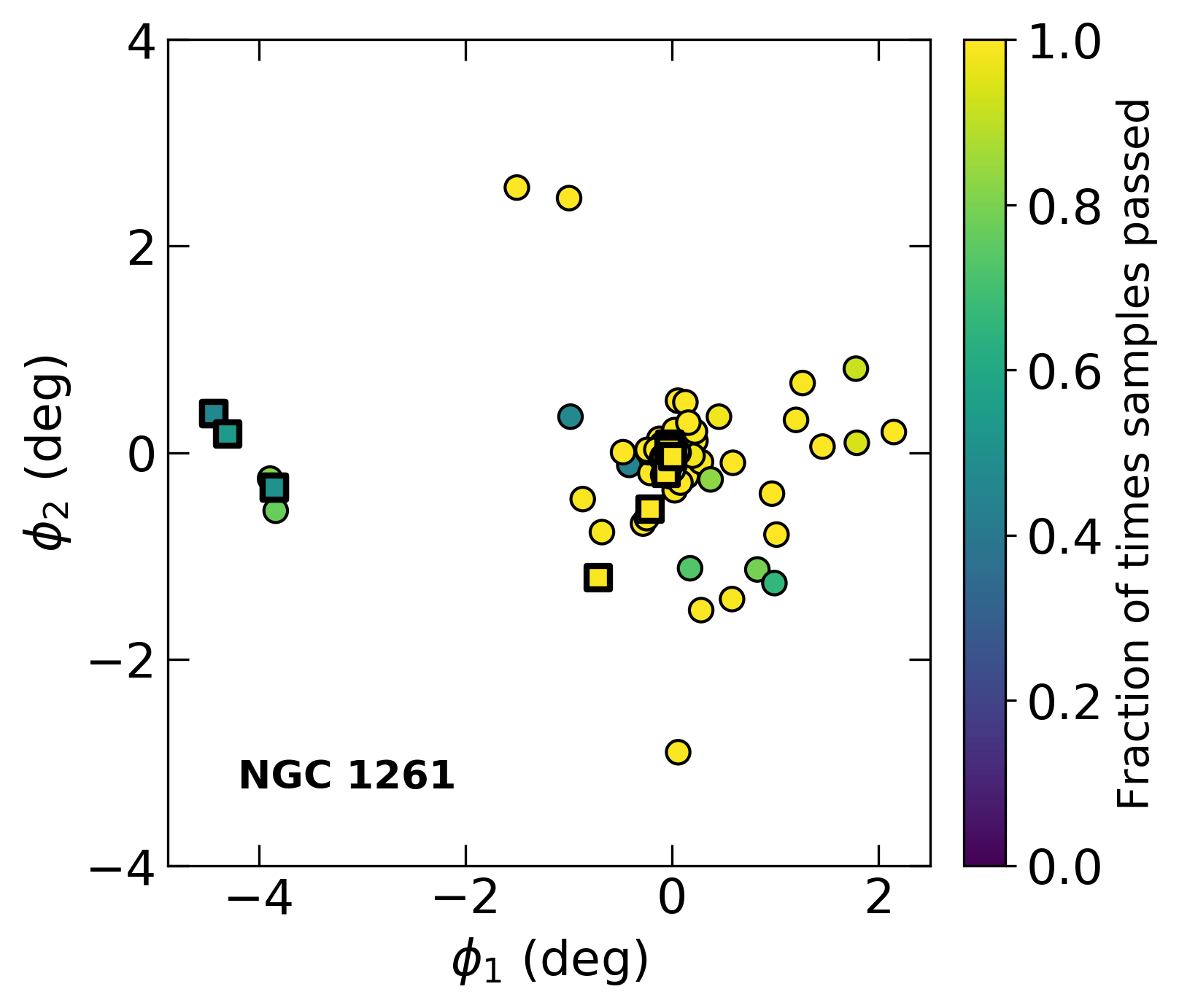} 
\includegraphics[width=\columnwidth]{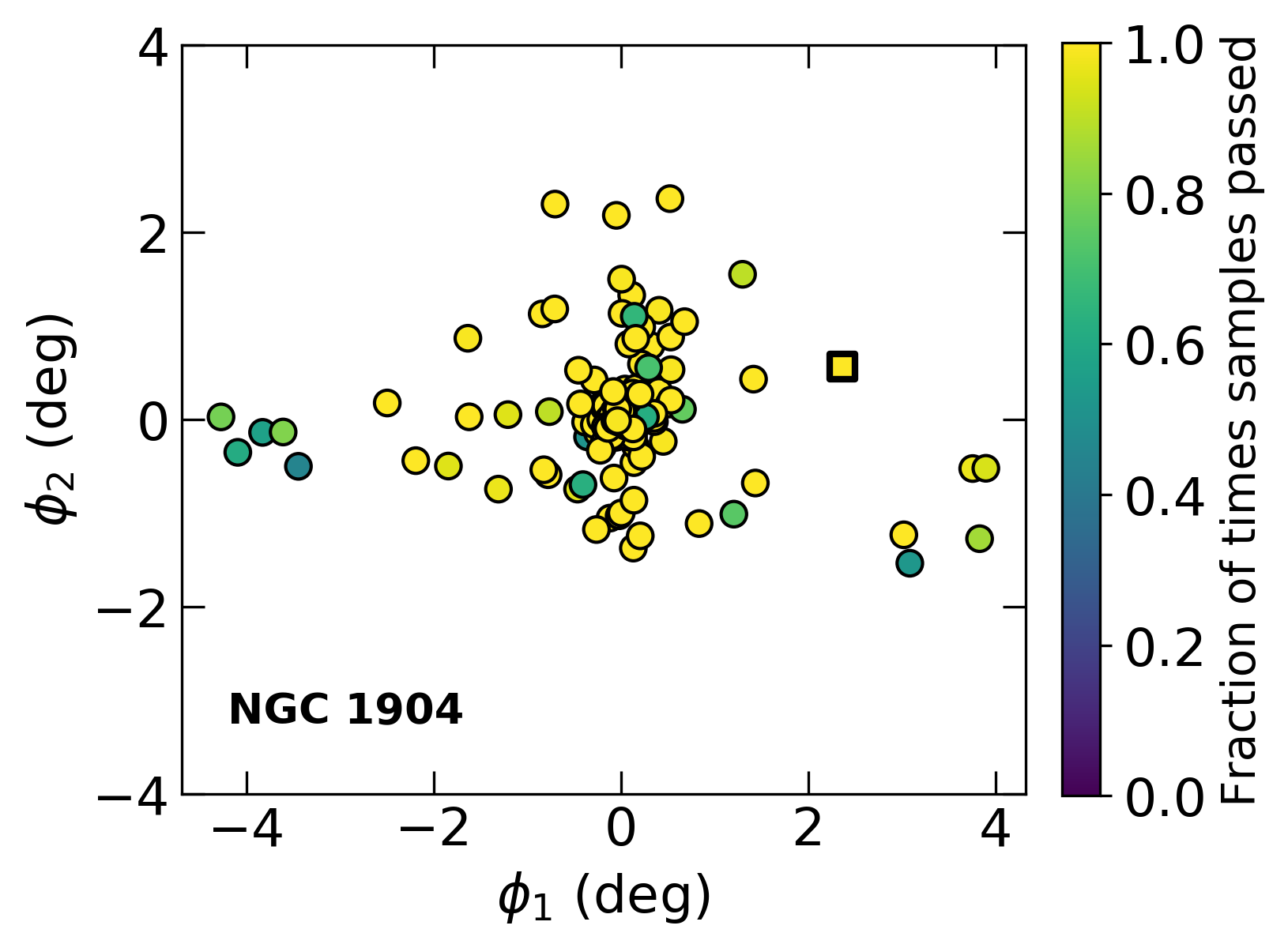} 
  \caption{We quantify the robustness of the probabilities we measure for each potential member by sampling 100 values from the posterior distributions of our mixture model parameters and checking how many times each star survives our cut using the model parameters from each of the 100 samples. Potential members in yellow have survived 100\% of the time while those closer to blue survived the corresponding fraction of times. Contaminants identified through their position in the CMD are indicated using a square marker.}
  \label{fig:robust}
\end{figure}

\subsection{Contamination and robustness of findings}
\label{sec:robustness}
In the Gaussian mixture modelling approach (Section~\ref{sec:methods}), we fit for a stream and a background component in the data. The properties of stars belonging to the stream component are assumed to vary along a polynomial as a function of its coordinate $\phi_1$, while stars belonging to the background have non-varying Gaussian distributions modelling their properties. A concern therefore could be that since we assume that a stream component exists, the algorithm will likely find a set of stars that have stream-like properties when the structures found are not necessarily real. In this case, the stream-like property would be the variation of the parameters modelling the extra-tidal GC stars along a thin polynomial distribution dependent on $\phi_1$. To address this concern we therefore repeat the mixture model runs we have performed while not fitting for one of the parameters available. In other words, we remove any information we have on one of our modelled parameters that shows a dependence on $\phi_1$, and repeat the runs with this information missing. 
Specifically, we fit for the distribution in $v_{gsr}$, $\mu_{\alpha}$ and $[\mathrm{Fe/H}]$ and take out any information about $\mu_{\delta}$. 
Hence, the 2-dimensional Gaussian modelling of the proper motion space of the GCs is now a 1-dimensional Gaussian modelling of the distribution in $\mu_{\alpha}$ alone. In this way the correlation between $\mu_{\alpha}$ and $\mu_{\delta}$ is also not considered, and therefore a large portion of the information that could constrain a stream component is removed. If the new high-probability member stars in these runs show a narrow distribution of $\mu_{\delta}$ even though no assumption is made on this parameter, this shows that the retrieved stars indeed follow a stream-like distribution that is not artificially invented. Additionally, if we retrieve the same high-probability members from the previous runs, this shows that our results are highly robust against missing information and large changes in the methodology.

\begin{figure*}[ht!]
  \centering
\includegraphics[width=\textwidth]{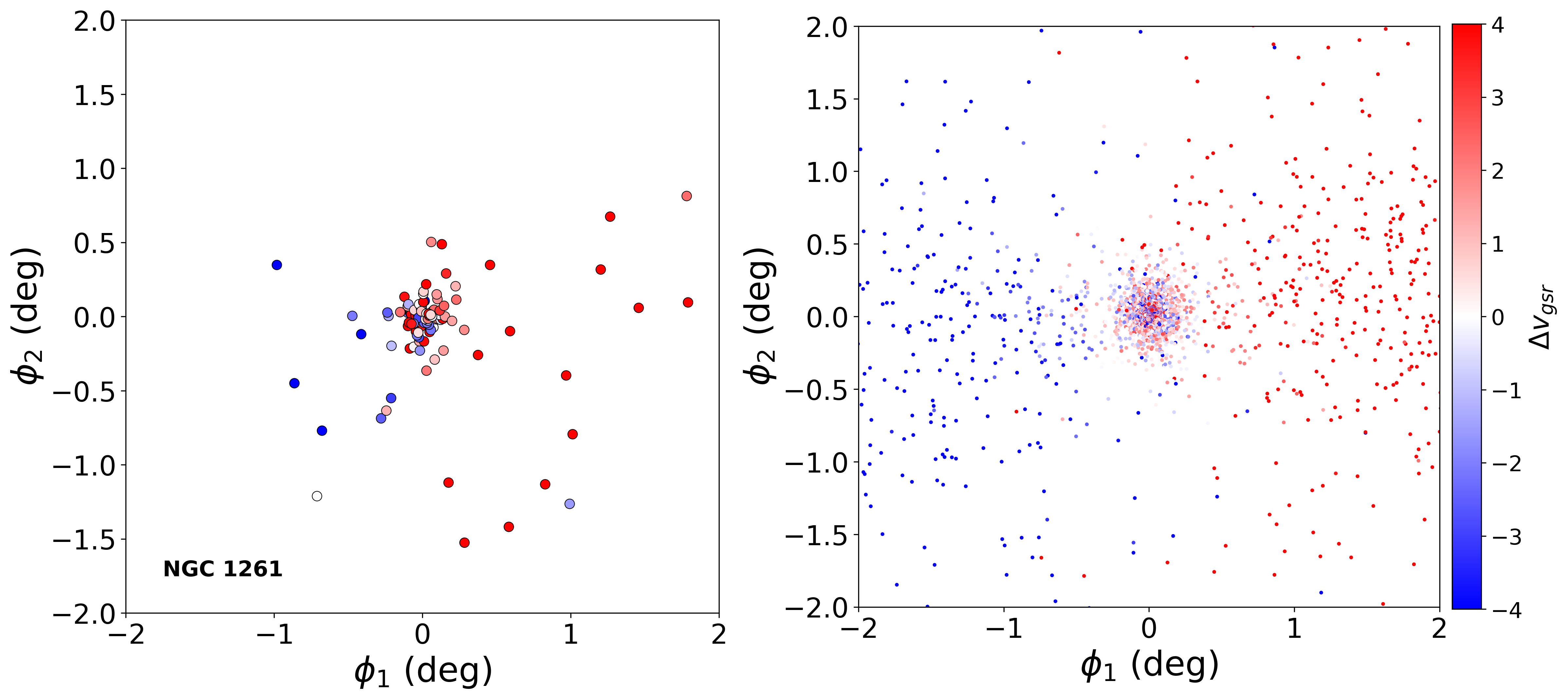} 
\includegraphics[width=\textwidth]{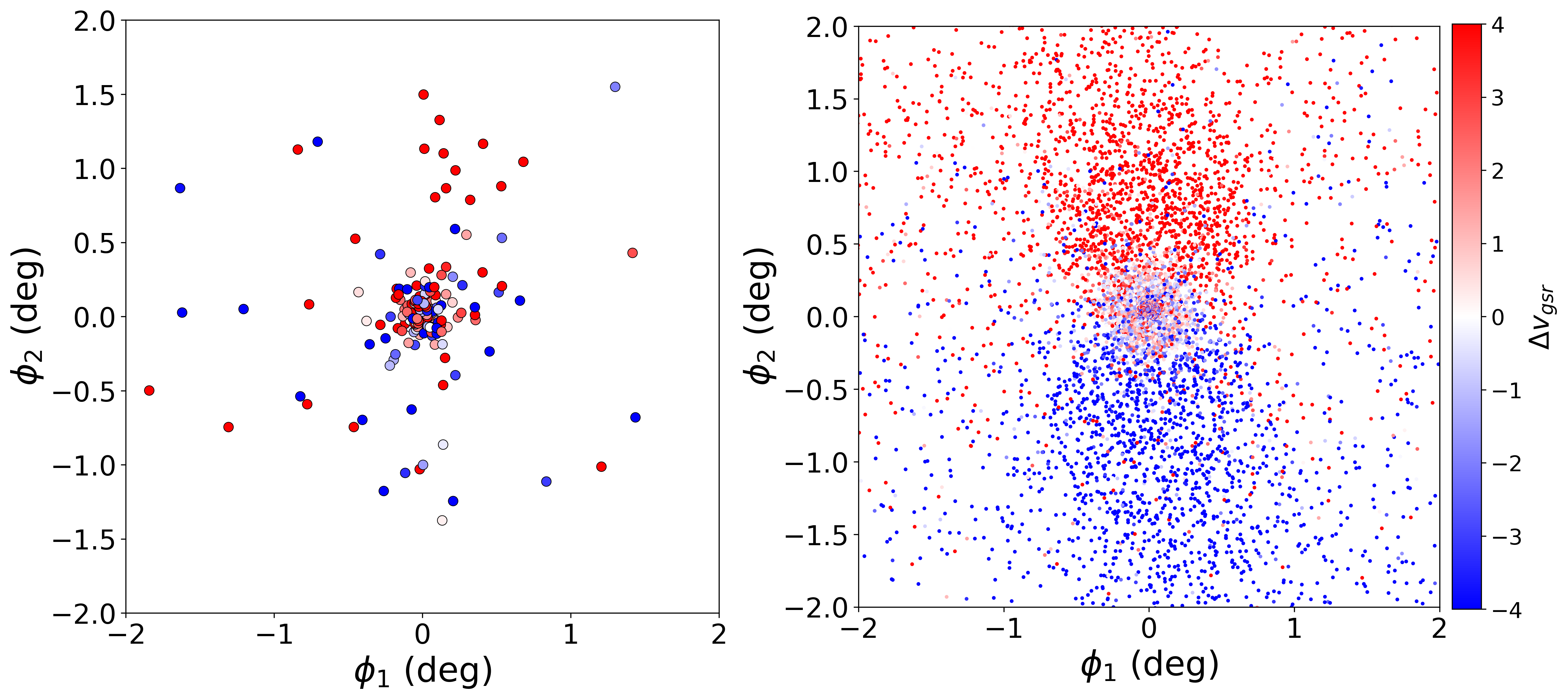} 
  \caption{Upper and lower rows refer to NGC~1261 and NGC~1904 respectively. Left panels: the potential member stars within  $\pm 2^{\circ}$ around the clusters colored by the radial velocity shifted with respect to the radial velocity of the cluster ($\Delta v_{gsr}$). Right panels: The same property is displayed using particles from the N-body simulation.}
  \label{fig:rotation}
\end{figure*}

The same range of priors assumed for the full-fit runs are used (refer to Table~\ref{tab:table1}).
The results from the new runs are displayed in Figure~\ref{fig:overlap} for NGC~1261 (left) and NGC~1904 (right). The best-fit tracks from the full-fit runs are shown by the dashed black line while the solid red line represents the best-fit tracks from the new runs excluding $\mu_{\delta}$. For both NGC~1261 and NGC~1904, we see great overlap between the two runs. We also extract the potential members from the new runs by enforcing a limit on the membership probability. Since by removing any information of $\mu_{\delta}$, we are using less information than when applying the full-fit runs, this increases the variation and the uncertainty of the outcomes. As a result, it is expected that the membership probability for originally high-probable member stars will decrease. Therefore, to extract potential members, we enforce a $50\%$ lower limit instead of the $60\%$ we used before. The stars that remain from this cut are shown in blue open circles in Figure~\ref{fig:overlap} and the common stars between these runs and the full-fit ones are shown using yellow open circles. As expected, since the runs excluding $\mu_{\delta}$ use less information, we find more stars that are more likely contaminants than when using the full-fit runs. These are the stars that were not common between the new and full-fit runs (blue points). However, the majority of potential members found using the full-fit runs have been retrieved by the new runs as well. This includes stars present in the extra-tidal structures found surrounding the two GCs. Most importantly, the stars that we retrieve with the less informative runs still follow the track found for $\mu_{\delta}$ even though no assumptions were made for this parameter. We repeat this same analysis now removing any information on $\mu_{\alpha}$ instead of $\mu_{\delta}$ and arrive at similar conclusions: we retrieve the potential members found using the full-fit runs though with expected larger contamination, and the retrieved stars follow a narrow distribution in $\mu_{\alpha}$ though no assumption was made on this parameter.

Regarding contamination, Figures~\ref{fig:CMDs1261} and \ref{fig:CMDs1904} have shown that some stars we retrieve do not align with the GCs' isochrones. These stars have been plotted in Figures~\ref{fig:results_1261} and \ref{fig:results_1904} to inspect their spatial distributions. We can also assess the contamination by taking parameter samples from the chains of the mixture model defined in Section~\ref{sec:methods}. We therefore sample 100 values of each mixture model parameter, and using each of these samples, the membership probability of the stars in our original selection is then calculated thus obtaining 100 probabilities for each star. The thresholds on the probability defined in Section~\ref{sec:results} are then applied and the fraction of samples in which each star passes the thresholding is counted. The results of this procedure are shown in Figure~\ref{fig:robust} for NGC~1261 (top) and NGC~1904 (bottom). The potential members for each GC are colored by the fraction of samples in which they had a membership probability greater than the threshold. Stars identified as contaminants given their position in the CMD are indicated with a square marker. We see that in the case of NGC~1261, one star with $\phi_1 < -2.5$ previously highlighted in red appears as potential members $\approx 80\%$ of the time, other stars closer to the cluster have a smaller fraction. The majority of the stars for this GC however, have robust membership probabilities. In the case of NGC~1904 (lower panel) we see that stars farthest away from the cluster in $\phi_1$ have probabilities that are not as robust as the other stars. These stars could therefore be contaminants. The majority of the stars forming the overdensities we detect on the top and bottom of this cluster however, have survived the thresholding $100\%$ of the time showing robust high probabilities against changes in the best fit model parameters. 

We therefore find that our method is detecting true structures surrounding the GCs which extend several degrees away from their centers. Additionally, although contamination is present, we find that it is minimal. This is also supported by previous work which have made note of the presence of these structures, primarily \citet{ShippEtal2018}, but have not studied them using proper motion or radial velocity and metallicity information.


\subsection{Understanding the tidal disruption of NGC 1261 and NGC 1904}

In order to better understand why NGC 1261 and NGC 1904 are tidally disrupting, we can compare their Jacobi radii to their present-day half-mass radius and the observed extent of the GC, which is parametrised by the tidal radius \citep{de_Boer+2019}. This is useful since the cluster is expected to strip heavily when the Jacobi radius is comparable to the half-mass radius. In addition, stars beyond the Jacobi radius are unbound so if the tidal radius is larger than the Jacobi radius, then there are stars to strip.

As calculated in Section~\ref{sec:methods}, NGC 1261's present-day Jacobi radius is $r_J = 170^{+4}_{-2}$~pc. This is much larger than its half-mass radius of 4.86~pc \citep{Baumgardt+2018} and its observed extent (i.e. tidal radius) of 51.51~pc \citep{de_Boer+2019} suggesting that NGC~1261 is currently in a relatively weak tidal field. However, at its pericenter of $0.6\pm0.1$~kpc, NGC 1261's Jacobi radius was $11^{+3}_{-2}$~pc: only $\sim 2.2$ half-mass radii, and significantly smaller than its present-day tidal radius. This suggests that stars in the outskirts of NGC 1261 would have stripped at its previous pericenter.

\begin{figure*}[ht!]
  \centering
\includegraphics[width=\textwidth]{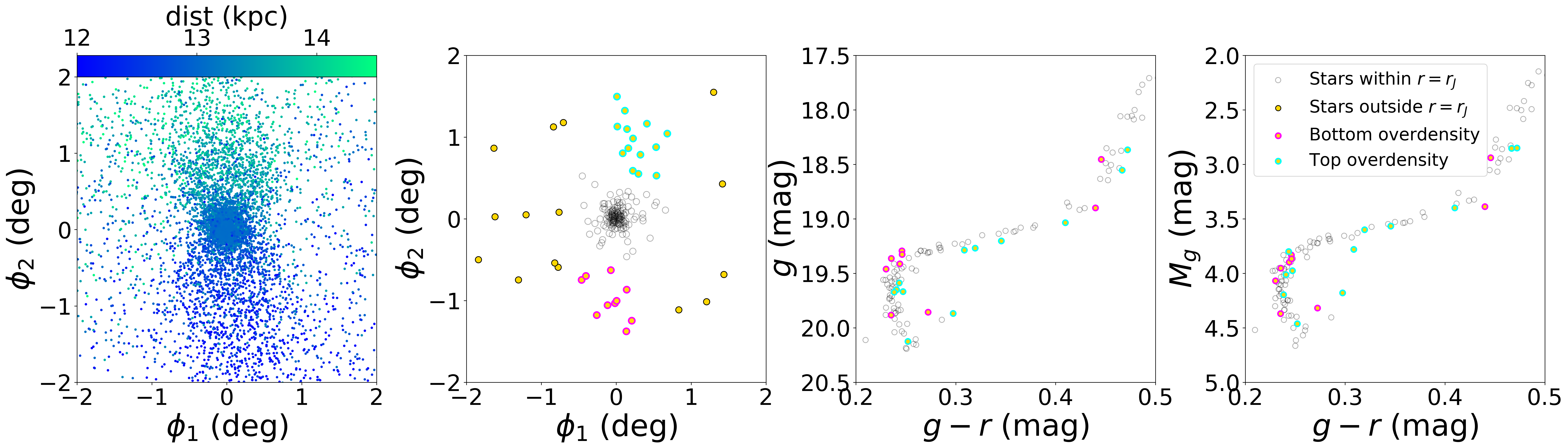} 
  \caption{Left: Distance gradient along $\phi_2$ seen in the simulations of NGC~1904. Middle left: the top and bottom overdensities of stars are selected in cyan and magenta respectively. Potential GC members that fall within $r_J$ are shown with empty black circles while those outside are shown in yellow. Middle right: the distance gradient can be seen in the main sequence and sub-giant branches given the observed differences in magnitude. Right: correcting for the distance differences leads to the overlapping of selected stars within the CMD supporting the present of the distance gradient. }
  \label{fig:Distance}
\end{figure*}

For NGC 1904, the present-day Jacobi radius is $r_J = 175^{+3}_{-3}$~pc. This is significantly larger than the half-mass radius of 4.33~pc \citep{Baumgardt+2018} and the tidal radius of 108.73~pc \citep{de_Boer+2019} suggesting that NGC 1904 is currently in a weak tidal field. However, at its pericenter of $0.12\pm0.06$~kpc, its Jacobi radius was $r_J = 3^{+1}_{-1}$~pc. This is only $\sim 0.7$ half-mass radii, suggesting that it experienced strong stripping at pericenter. Indeed, the dipole-like morphology of the tidal debris seen in the top and bottom right panels of Figure~\ref{fig:results_1904} was likely stripped at the previous pericenter. Thus, although both GCs are currently in a weak tidal field, they both experienced much stronger tides at pericenter.

\subsection{Further insights into the extra-tidal features of the GCs}

In Section~\ref{sec:introduction}, we introduced the physical mechanism behind which tidal tails around GCs form and the characteristic S-shaped morphology that they take. When a GC is not at pericenter, the inner-most parts of the tails point towards the Galactic center, with the the outer parts of the tails (at larger distances from the cluster center) moving towards the orbital path  \citep{MontuoriEtal2007, KlimentowskiEtal2009}. 
In Figure~\ref{fig:rotation}, we study this behaviour in more detail by looking at the radial velocity of each star with respect to the radial velocity of the cluster $\Delta v_{gsr} = v_{gsr} - v_{gsr}^{GC}$, where $v_{gsr}^{GC}$ is the mean radial velocity of each GC taken from \citep{Baumgardt+2018} and transformed to the galactic standard of rest.  
Figure~\ref{fig:rotation} shows the distribution of stars within $\pm 2^\circ$ around NGC~1261 (top row) and NGC~1904 (bottom row) for our potential members (left) and simulations (right). Each star is colored by its value of $\Delta v_{gsr}$. For NGC~1261, the simulations predict a gradient in $\Delta v_{gsr}$ along $\phi_1$, which is also visible using the observed sample from the corresponding left panel. For NGC~1904, we observe a clear change in the direction of $\Delta v_{gsr}$ for the inner tidal tails evidenced by the switching of the sign of $\Delta v_{gsr}$ between $\phi_2<0$ and $\phi_2>0$. This behaviour is compatible between the observational sample (left) and the simulations (right). This therefore shows that the top and bottom overdensities of stars (in both the observations and simulations) are moving in opposite directions, pointing towards the apparent rotational motion undergone by these stars as they escape the cluster's potential \citep{MontuoriEtal2007}. 

Further inspection of the top and bottom overdensities of stars can be performed by looking at their corresponding distances. In the left-most panel of Figure~\ref{fig:Distance}, we color the particles in the N-body simulation with their heliocentric distances in the same area on the sky as shown in Figure~\ref{fig:rotation}. We observe an expected distance gradient along $\phi_2$ with $\approx 2$~kpc difference between stars in the top and bottom overdensities. To check whether this gradient is also detected within our sample, we select the stars in the top (cyan) and bottom (magenta) overdensities, as well as the stars within $r_J$ (grey) as shown in the middle left panel of the same figure. We then plot the CMD of these stars in the middle right panel zooming in on the main sequence and sub-giant branches. We observe a difference in magnitude detected as a vertical offset in this panel between the three selected groups of stars. 
To further confirm this distance gradient, we attribute a distance measure to each star in our sample equal to the distance of its nearest neighbor in the simulation particles. By correcting the magnitudes of the potential members given their distances, we then obtain a measure of their absolute magnitudes $M_g$. The right panel of the figure shows the result of this procedure where we observe that the three selected groups of stars now overlap after this correction, supporting the presence of the distance gradient predicted by the simulations and further consolidating the true membership of these stars to the GC.

Additionally, there exists a strong correlation between the orientation of inner tidal tails and the position of the GC along the orbit. For near-circular orbits, the angular velocity is constant and therefore results in an almost constant orientation of the tails with respect to the direction of the cluster orbit and the Galactic center. For eccentric orbits, however, the behaviour becomes more complicated. When approaching apocenter, stars between the cluster and the Galactic center reach their apocenters before the GC does, decelerating, being on tighter orbits than the rest of the tail. On the other hand, stars between the cluster and the Galactic anti-center will not have yet reached their apocenters and so will be moving faster. The resulting effect seen is that near apocenters, the inner tails are oriented along the direction to the Milky Way center. When approaching pericenter, stars between the GC and the Galactic center speed up with respect to the GC, again being on tighter orbits, while those on the opposite side of the GC slow down. This leads the tails to become elongated along the direction of the GC orbit \citep{MontuoriEtal2007, KlimentowskiEtal2009, KupperEtal2012}. 

\begin{figure*}[ht!]
  \centering
\includegraphics[width=\textwidth]{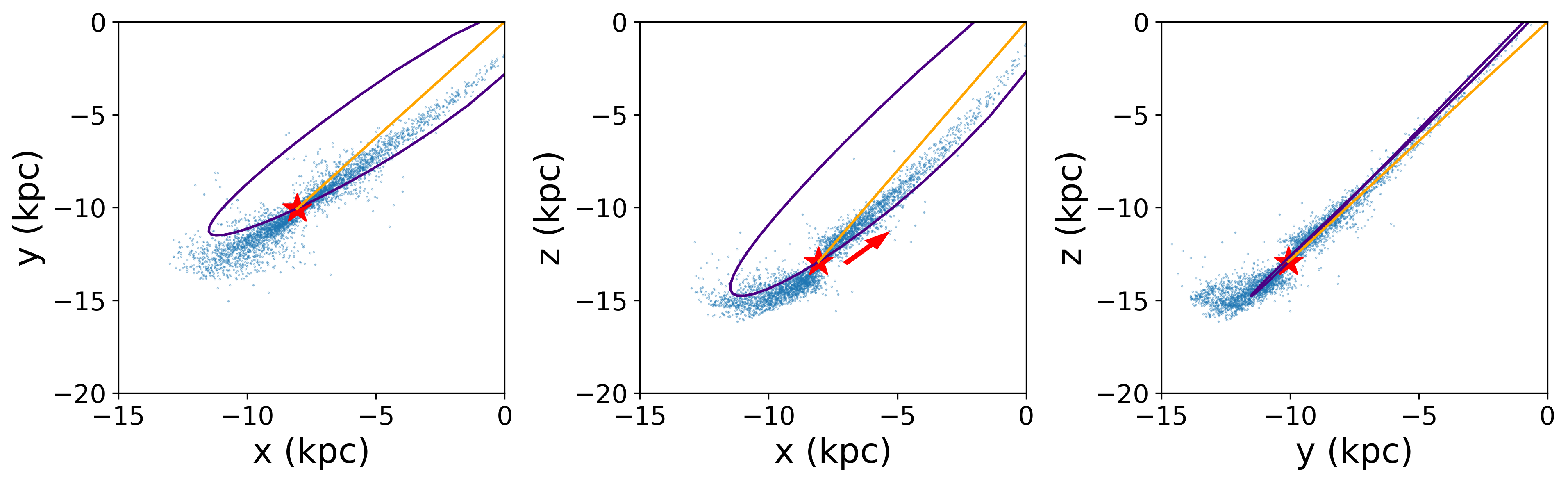} 
\includegraphics[width=\textwidth]{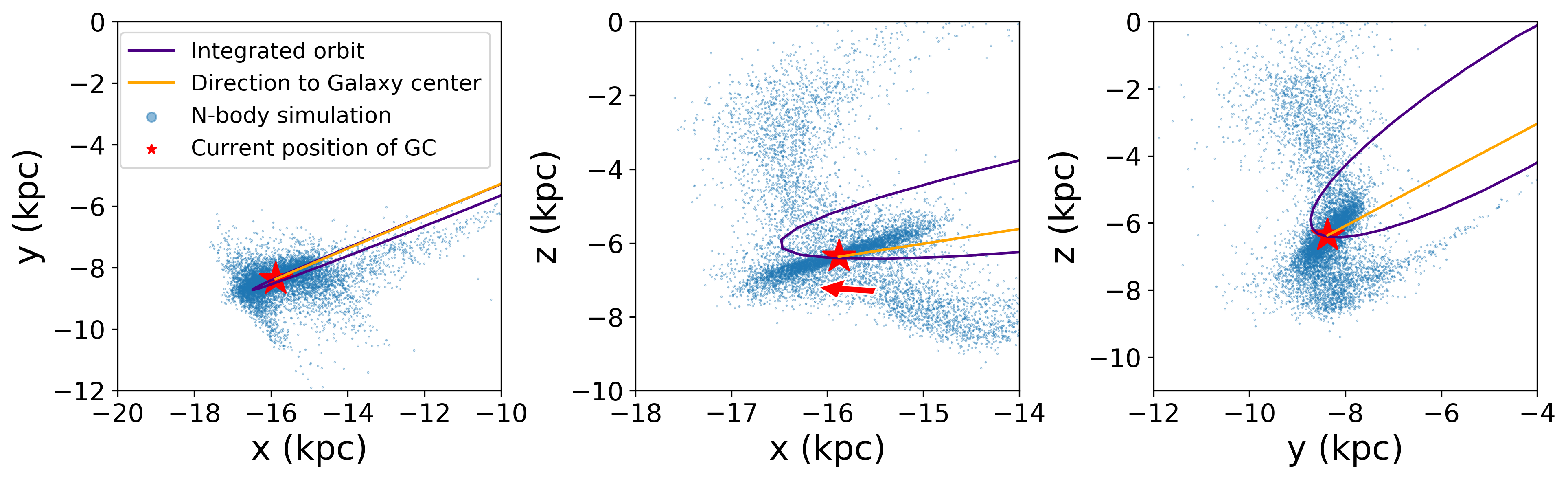} 
  \caption{Orbits of NGC~1261 (upper row) and NGC~1904 (lower row). Each column presents a different plane projection in galactocentric coordinates. The red star indicates the present position of each GC and the simulation particles are over-plotted in this coordinate system. The purple line represents the integrated orbit of each cluster where the direction of orbit is indicated by the red arrow in the middle panels. The orange line connects the cluster to the center of the Galaxy. We see that NGC~1261 is close to apocenter but has passed it and is moving closer to pericenter while NGC~1904 is approaching and is very close to its orbit's apocenter.}
  \label{fig:orbits}
\end{figure*}

This dynamic can be probed by inspecting the orientation of tails with respect to the current phases of the GCs along the orbit.  In Figure~\ref{fig:orbits}, we plot in red and using Galactocentric coordinates the current position of each GC along its integrated orbit plotted in purple. The upper row refers to NGC~1261 and the bottom row to NGC~1904. Each column is a projection on one of the planes defined in this coordinate system. The orbit has been integrated using the gravitational potential defined in Section~\ref{sec:simData}. Arrows in the middle panels show the direction in which the clusters are moving. The orange line connects the cluster to the center of the Galaxy, and the respective N-body simulation scatter is also plotted in blue. From Figure~\ref{fig:orbits}, we can see that NGC~1261 has recently passed apocenter and is heading towards pericenter, while NGC~1904 is close to reaching apocenter. The difference in the tidal tails can then clearly be seen as explained by the theory: the inner tails for NGC~1261 point towards the orbit while in the case of NGC~1904, we see the dipole morphology of the inner tails pointing in the direction of the Milky Way center. This shows that the overdensities we detect on the top and bottom of NGC~1904 are a result of this cluster approaching apocenter and the reason we do not see these overdensities in the case of NGC~1261 is because this GC is in a different phase in its orbit where it has passed its apocenter and is now heading towards pericenter\footnote{See \url{https://www.youtube.com/watch?v=lTUK1mfpM70} and \url{https://www.youtube.com/watch?v=8bHzEqXmwZU} for movies showing the 3d view of NGC 1904 and NGC 1261 respectively.}. Finally, we also note that the Sun's location relative to the orbital plane of NGC 1904 is also crucial for being able to see the radially extended debris on the sky. Since the Sun sits $\sim$4~kpc above the orbital plane of NGC 1904, we can see this radial extension projected onto the sky instead of only being able to see it with precise distances.

\begin{figure}[ht!]
  \centering
\includegraphics[width=\columnwidth]{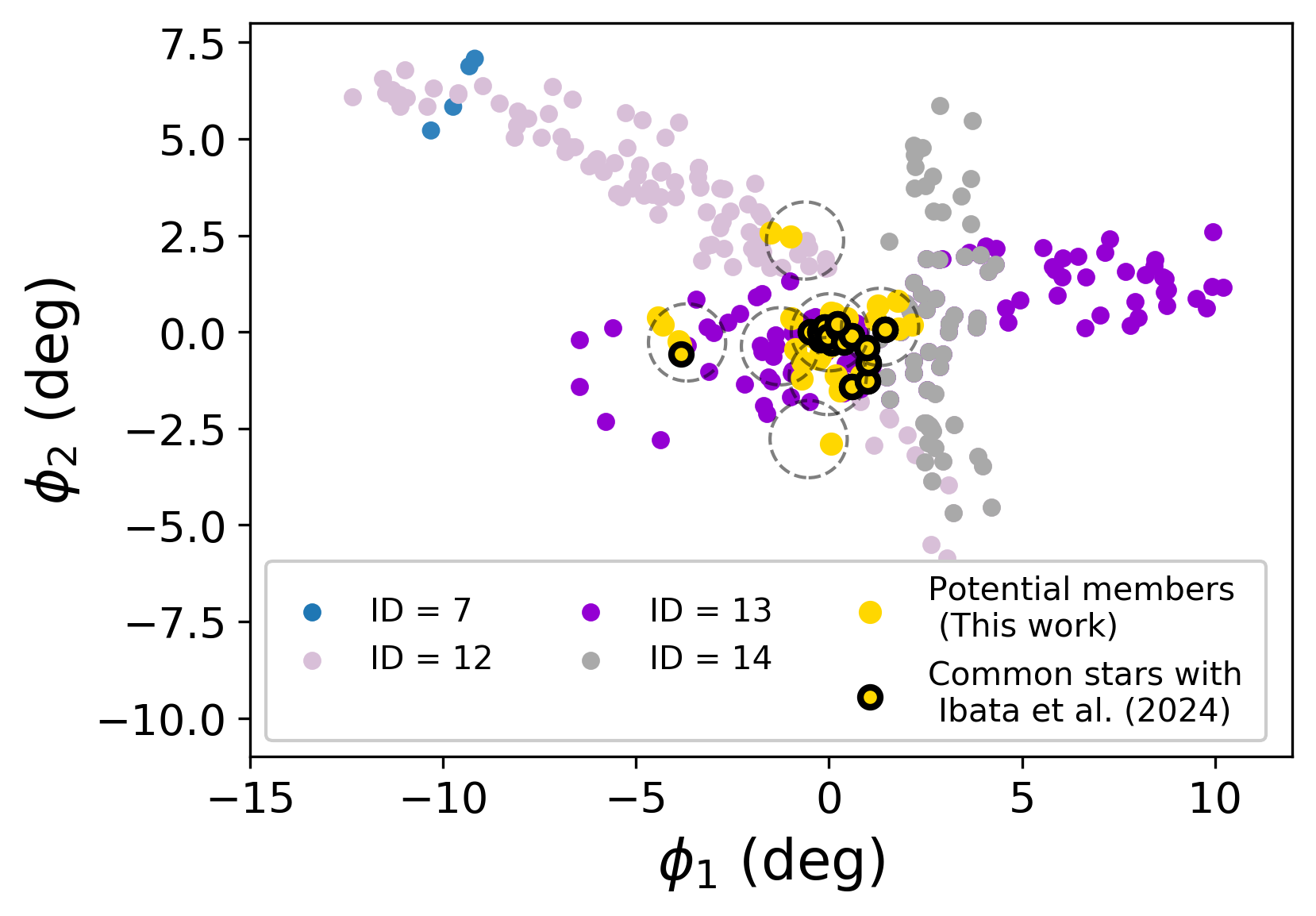} 
  \caption{Comparison with the results of \citet{IbataEtal2023} around NGC~1261. Streams found by \citet{IbataEtal2023} surrounding NGC~1261 and thought to be associated to the GC are plotted here colored differently according to the corresponding stream ID. We overplot the potential members for NGC~1261 found in this paper in yellow and outline the common ones between the two works in black. We also show the areas inspected by $S^5$ (dashed circles).}
  \label{fig:Comaprison}
\end{figure}

We also compare our potential members detected for NGC~1261 with the results presented in \citet{IbataEtal2023}. In the latter work, two streams were speculated to be related to NGC~1261 given that they share the same position on the sky. The streams were labelled NGC~1261a and NGC~1261b. In Figure~\ref{fig:Comaprison}, we plot the stellar streams detected in \citet{IbataEtal2023} that surround NGC~1261, where we color each star in the figure according to its corresponding stream index. We observe that at least two streams crisscross this area. We also overplot in yellow the potential members detected in the current work, with the stars common with \citet{IbataEtal2023} outlined in black. The fields surveyed by $S^5$ are indicated by the dashed circles. In total, we find 34 common stars, 20 of which are within $r_J$ and 14 are outside. From this figure we see the extent of the extra-tidal features around NGC~1261 which goes beyond the fields probed by $S^5$. This also supports the presence of a wide distribution of member stars surrounding this GC. We note that for NGC~1261, our simulations do not reproduce the double stream morphology inspected in this work and seen in \citet{ShippEtal2018} and \citet{IbataEtal2023}. This therefore points towards more complex dynamics that serve to form this second stream that should be incorporated in the simulations to retrieve the structures observed in these several works. One possible explanation of this complex morphology is the Milky Way bar. Previous works have found that the rotating bar can have a significant effect on stellar streams \citep[e.g.][]{Hattori+2016,Price-Whelan+2016,Erkal+2017,Pearson+2017}, and can create complex morphology in the tidal debris close to the cluster \citep[e.g.][]{Dillamore+2024}. Given that the orbit of NGC~1261 is highly eccentric with a small pericenter, this makes the interaction of this GC with the Milky Way bar very possible and also frequent given its small orbital period. Additionally, NGC~1261 is known to have undergone more than 10 orbital laps around the Galaxy in the past 2 Gyr \citep{WanEtal2023} similarly due to its short orbital period. At each pericenter passage, the GC sheds some its stars producing a new stream that follows its orbit as a result. These streams can overlap on the long run and produce "multiple tidal tail" features around GCs, and could be the case explaining the extra-tidal features seen for NGC~1261.

\subsection{GC properties}
\label{sec:properties}

Given that we fit the mean and intrinsic dispersions for the properties of the GCs while fitting for the stream and background components, we can compare our measured values to previous literature estimates. For NGC~1261, we measure a radial velocity dispersion of $\sigma_v = 2.82 \pm 0.32$~${\rm km~s}^{-1}$ which is in agreement with \citet{WanEtal2023} within 2 standard deviations. For NGC~1904, we measure $\sigma_v = 3.06 \pm 0.30$~${\rm km~s}^{-1}$ which is smaller than that measured by \citet{WanEtal2023} of $7.52^{+2.18}_{-1.51} $~${\rm km~s}^{-1}$, though in the latter work it is mentioned that the last two stars in their radial bins contribute the most to the dispersion, and when excluded from their sample, the dispersion drops to $2.45^{+1.08}_{-0.81} $~${\rm km~s}^{-1}$ which then becomes in good agreement with our estimate. Note that \citet{WanEtal2023} consider an area of $1^{\circ}$ around their sample of GCs which is $\approx r_J$ in the case of NGC~1904.

In terms of the mean metallicity, we measure  $\overline{[\mathrm{Fe/H}]} = -1.29 \pm 0.02$ for NGC~1261 and $\overline{[\mathrm{Fe/H}]} = -1.61 \pm 0.02$ for NGC~1904. These estimates are in excellent agreement with works such as \citet[][2010 edition]{Harris1996}, \citet{FerraroEtal1999}, \citet{MunozEtal2021}, and \citet{WanEtal2023}. As for the metallicity dispersion, we measure $\sigma_{\mathrm{[Fe/H]}} = 0.19 \pm 0.02$ for NGC~1261 and $\sigma_{\mathrm{[Fe/H]}} = 0.21 \pm 0.02$ for NGC~1904. These values are larger than what is presented in works such as \citet{MunozEtal2021}, \citet{LimbergEtal2022}, and \citet{WanEtal2023} which can be an indication of high contamination in our sample or an underestimation of the measurement errors. We argue here that the issue is an underestimation of measurement uncertainties.

To gain better understanding of the source of the issue, we split the detected potential members i.e. stars that have passed the probability threshold, between stars within $r_J$ and those outside and repeat the calculation of the intrinsic metallicity dispersion on both these groups. We also exclude the stars that have been deemed as contaminants given their positions in CMD i.e. stars indicated throughout this work with a square marker. The calculation is performed by fitting a Gaussian function to the distribution of metallicities of either group following an MCMC framework using the {\ttfamily emcee} package with 100 walkers and 1000 steps. Specifically, we fit for the mean metallicity and intrinsic metallicity dispersion. The same analysis is performed for the velocity dispersion. In other words, we calculate the velocity dispersion of the stars within and outside $r_J$, by fitting a Gaussian function to the distribution of $v_{gsr} - v(x)$, where we remind $v(x)$ is the best fit polynomial for the variation of $v_{gsr}$ against $\phi_1$.

\begin{table*}[t]
\caption{Recalculated metallicity and velocity dispersions for stars inside and outside $r_J$ of both NGC~1261 and NGC~1904.}
\centering          
\begin{tabularx}{\textwidth}{@{\extracolsep{\fill}}
ccccccccc@{}
} 
\toprule
\thead{ }
&N (stars)
&$\sigma_{\mathrm{[Fe/H]}}$
&$\sigma_v~({\rm km~s}^{-1})$
&N (star)
&$\sigma_{\mathrm{[Fe/H]}}$
&$\sigma_v~({\rm km~s}^{-1})$ 

\\
\thead{ }
& NGC~1261
& NGC~1261
& NGC~1261
& NGC~1904
& NGC~1904
& NGC~1904
\\
\midrule

$r < r_J$  
& 88
& $0.13^{+0.02}_{-0.02}$ 
& $2.55^{+0.33}_{-0.28}$ 
& 143
& $0.19^{+0.02}_{-0.02}$ 
& $2.49^{+0.25}_{-0.23}$ \\
\midrule

$r > r_J$ 
& 28
& $0.15^{+0.05}_{-0.04}$ 
& $3.00^{+1.05}_{-0.79}$ 
& 51
& $0.13^{+0.04}_{-0.03}$ 
& $4.56^{+1.02}_{-0.84}$ \\

\bottomrule
\end{tabularx}
\label{tab:table2}
\end{table*}

The results are presented in Table~\ref{tab:table2} for both GCs where the total number of stars in each region is also indicated. For NGC~1261, we observe that the metallicity dispersions for stars inside and outside $r_J$ are compatible. This indicates low contamination in the stream component and instead points towards an underestimation of the measurement uncertainties. Additionally, the dispersions are now smaller and in agreement with previous work now that a threshold on the probability has been applied. Similarly, we retrieve comparable velocity dispersions for stars inside and outside $r_J$ which remain consistent with values present in the literature with an slight increase from $2.55^{+0.33}_{-0.28}$~${\rm km~s}^{-1}$  to $3.00^{+1.05}_{-0.79}$~${\rm km~s}^{-1}$ which is expected as we move radially away from the cluster center. For NGC~1904, $\sigma_{\mathrm{[Fe/H]}}$ inside $r_J$ stayed the same as when fitting for the entire set-up $(0.19^{+0.02}_{-0.02})$. Given the low level of contamination in this region especially evidenced by the CMD distribution in Figure~\ref{fig:CMDs1904}, we expect that the measurement errors for the stars inside $r_J$ are underestimated leading to the inflated dispersion. We thus suspect that the {\ttfamily rvspecfit}  metallicities have a systematic accuracy floor of $\sim 0.2$ dex, but might be the best that can be done with a low-resolution spectrograph like AAOmega. For stars, outside $r_J$ however, we see that the metallicity dispersion drops to $0.13^{+0.04}_{-0.03}$ which is now consistent with estimates provided in \citet{WanEtal2023} and therefore consistent with an unresolved dispersion. For the velocity dispersion we measure $2.49^{+0.25}_{-0.23}$~${\rm km~s}^{-1}$ which is still in agreement with the estimate of \citet{LimbergEtal2022} and \citet{WanEtal2023} when the latter exclude the outermost two stars in their analysis of this cluster. For stars outside $r_J$, we measure a radial velocity dispersion of $4.56^{+1.02}_{-0.84}$~${\rm km~s}^{-1}$, an expected increase which then is in agreement with the estimate of \citet{WanEtal2023} within the given error margins.

\section{Conclusions}
\label{sec:conclusion}

In this work, we have explored the extra-tidal features of the globular clusters (GCs) NGC~1261 and NGC~1904. 
Given the interesting morphology of the features surrounding the two GCs, the Southern Stellar Stream Spectroscopic Survey ($S^5$) performed follow-up observations of fields surrounding the clusters where radial velocity and metallicity measurements were gathered for the target stars. In this work, we have used these observations to gather new information about the structures surrounding the two GCs and linked their presence to their dynamical properties. The analysis performed and results obtained can be summarized as follows: 

\begin{enumerate}
    \item With the use of information on the proper motions, radial velocity and metallicity of the targeted stars, we applied a Bayesian mixture modelling approach to provide a membership probability thereby separating high-probability stars from others that are more likely field stars. 
    \item We identified potential member stars associated with these two GCs within an area spanning 10 times their Jacobi radius. By inspecting the spatial distribution of the potential members, we confirmed the results of \citet{ShippEtal2018}. We observed stars distributed along the direction of the clusters' orbits as well as distributions of stars not aligned with this direction and forming a cross-shaped pattern with the former in the case of NGC~1904 and a broad distribution of stars in the case of NGC~1261. 
    \item We performed N-body simulations of the two GCs and compared what is expected from the simulations with regards to the evolution of the properties of the escaped cluster stars along the orbit and the properties of the detected potential members from our sample. It is clear that there is good agreement between the observational sample and the simulation predictions for NGC~1904 but a clear comment cannot be made with respect to NGC~1261. 
    \item Using DECam photometry, we inspected the color-magnitude diagrams of the potential member stars and confirmed that the majority of them lie along an isochrone-like distribution which is expected for GCs. This final check is a powerful independent confirmation that we are correctly identifying members and the structures in which they are now found. 
    \item We discussed the origin of the observed structures and linked their presence to the phase of orbit of each GC. Given that NGC~1904 is approaching apocenter, we expect a clear distinction between inner and outer tidal tails, where the inner parts of its tidal tails are expected to point towards the Milky Way center. On the other hand, NGC~1261 has recently passed its apocenter passage and is now on its way to pericenter. Therefore, the inner parts of its tidal tails are expected to have closer alignment with its orbit instead. We discussed that the remaining stream for this cluster can be associated instead with the fact that NGC~1261 has undergone several orbits around the galaxy and its possible interaction with the Milky Way bar.
\end{enumerate}

Similar to NGC~1261 and NGC~1904, other Milky Way GCs exhibit signs of multiple tails such as NGC~288 and NGC~2298. Therefore, it is possible to perform the analysis presented here for these systems as well. 
Additionally, given that the studied GCs are found close to their apocenter passages, the epicyclic motion within their tidal tails would be theoretically prominent and so one can attempt to look for periodicity in the distribution and dynamics of the stars in the tails. Moreover, follow-up high-resolution spectroscopic measurements of the potential members found in this work could help confirm membership and improve the precision of our estimates. We leave such explorations for future work. 

\section*{Data availability}
The full versions of Tables~\ref{table:ngc1261} and \ref{table:ngc1904} are only available in electronic form at the CDS via anonymous ftp to \hyperlink{cdsarc.u-strasbg.fr (130.79.128.5)}{cdsarc.u-strasbg.fr (130.79.128.5)} or via \hyperlink{http://cdsweb.u-strasbg.fr/cgi-bin/qcat?J/A+A/}{http://cdsweb.u-strasbg.fr/cgi-bin/qcat?J/A+A/}.

\begin{acknowledgements}
This work is supported by the DSSC Doctoral Training Program of the University of Groningen. T.S.L. acknowledges financial support from Natural Sciences and Engineering Research Council of Canada (NSERC) through grant RGPIN-2022-04794. DBZ, GFL and SLM acknowledge support from the Australian Research Council through Discovery Program grant DP220102254.
 SK acknowledges support from the Science \& Technology Facilities Council (STFC) grant ST/Y001001/1.

This paper includes data obtained with the Anglo-Australian Telescope in Australia. We acknowledge the traditional owners of the land on which the AAT stands, the Gamilaraay people, and pay our respects to elders past and present.

This work was supported by the Australian Research Council Centre of Excellence for All Sky Astrophysics in 3 Dimensions (ASTRO 3D), through project number CE170100013.

For the purpose of open access, the author has applied a Creative Commons Attribution (CC BY) licence to any Author Accepted Manuscript version arising from this submission.

This work presents results from the European Space Agency (ESA) space
mission Gaia. Gaia data are being processed by the Gaia Data
Processing and Analysis Consortium (DPAC). Funding for the DPAC is
provided by national institutions, in particular the institutions
participating in the Gaia MultiLateral Agreement (MLA). The Gaia
mission website is https://www.cosmos.esa.int/gaia. The Gaia archive
website is https://archives.esac.esa.int/gaia.

This project used public archival data from the Dark Energy Survey
(DES). Funding for the DES Projects has been provided by the
U.S. Department of Energy, the U.S. National Science Foundation, the
Ministry of Science and Education of Spain, the Science and Technology
Facilities Council of the United Kingdom, the Higher Education Funding
Council for England, the National Center for Supercomputing
Applications at the University of Illinois at Urbana-Champaign, the
Kavli Institute of Cosmological Physics at the University of Chicago,
the Center for Cosmology and Astro-Particle Physics at the Ohio State
University, the Mitchell Institute for Fundamental Physics and
Astronomy at Texas A\&M University, Financiadora de Estudos e
Projetos, Funda{\c c}{\~a}o Carlos Chagas Filho de Amparo {\`a}
Pesquisa do Estado do Rio de Janeiro, Conselho Nacional de
Desenvolvimento Cient{\'i}fico e Tecnol{\'o}gico and the
Minist{\'e}rio da Ci{\^e}ncia, Tecnologia e Inova{\c c}{\~a}o, the
Deutsche Forschungsgemeinschaft, and the Collaborating Institutions in
the Dark Energy Survey.  The Collaborating Institutions are Argonne
National Laboratory, the University of California at Santa Cruz, the
University of Cambridge, Centro de Investigaciones Energ{\'e}ticas,
Medioambientales y Tecnol{\'o}gicas-Madrid, the University of Chicago,
University College London, the DES-Brazil Consortium, the University
of Edinburgh, the Eidgen{\"o}ssische Technische Hochschule (ETH)
Z{\"u}rich, Fermi National Accelerator Laboratory, the University of
Illinois at Urbana-Champaign, the Institut de Ci{\`e}ncies de l'Espai
(IEEC/CSIC), the Institut de F{\'i}sica d'Altes Energies, Lawrence
Berkeley National Laboratory, the Ludwig-Maximilians Universit{\"a}t
M{\"u}nchen and the associated Excellence Cluster Universe, the
University of Michigan, the National Optical Astronomy Observatory,
the University of Nottingham, The Ohio State University, the OzDES
Membership Consortium, the University of Pennsylvania, the University
of Portsmouth, SLAC National Accelerator Laboratory, Stanford
University, the University of Sussex, and Texas A\&M University.
Based in part on observations at Cerro Tololo Inter-American
Observatory, National Optical Astronomy Observatory, which is operated
by the Association of Universities for Research in Astronomy (AURA)
under a cooperative agreement with the National Science Foundation.

\end{acknowledgements}

%
%
\bibliographystyle{aa}
\bibliography{references}

\appendix

\onecolumn

\section{Rotation matrices}
\label{appendix}

\begin{equation*}
\centering
\begin{matrix}R_{1261} = 
\begin{pmatrix}

0.38122504 & 0.42440515 & -0.82130855 \\
0.10333836 & 0.8632682 & 0.49405384 \\
0.91868855 & -0.27321838 & 0.28524212 
\end{pmatrix}    
\end{matrix}
\end{equation*}

\begin{equation*}
\centering
\begin{matrix}R_{1904} = 
\begin{pmatrix}
 0.14163201 & 0.89869552 & -0.41507437 \\
-0.68463391 & 0.39177768 & 0.61464352 \\
0.71499425  & 0.19712079 & 0.67076569

\end{pmatrix} 
\end{matrix}
\end{equation*}

\section{Potential members}
\label{appendixB}

\begin{table*}[h!]
\small
\caption{List of potential members of NGC~1261. Only 10 members are shown here while the full list is available at the CDS.}             
\label{table:ngc1261}      
\centering      
\renewcommand{\arraystretch}{1.25} 
\begin{tabular}{@{}l@{ }c@{ }c@{ }c@{ }c@{ }c@{ }c@{ }c@{ }c@{ }c@{ }c@{ }c@{ }c@{ }c@{ } c@{ }
}     
\hline 
\multicolumn{1}{p{2.5cm}}{\centering Gaia ID \\ (Gaia DR3)}
&\multicolumn{1}{p{0.9cm}}{\centering RA \\(ICRS)}
&\multicolumn{1}{p{0.9cm}}{\centering Dec \\(ICRS)}
&\multicolumn{1}{p{1.2cm}}{\centering $\mu_{\alpha}$ \\ (mas/ yr)}
&\multicolumn{1}{p{1.2cm}}{\centering $\mu_{\delta}$ \\ (mas/ yr)}
&\multicolumn{1}{p{1.4cm}}{\centering $(BP-RP)_0$ (mag)}
&\multicolumn{1}{p{0.75cm}}{\centering $G_0$ (mag)}
&\multicolumn{1}{p{0.75cm}}{\centering $v_{rad}$ \\ (km/s) }
&\multicolumn{1}{p{1cm}}{\centering $\sigma_{meas}^{v}$ \\ (km/s)}
&\multicolumn{1}{c
}{\centering [Fe/H]}
&\multicolumn{1}{p{1cm}}{\centering $\sigma_{meas}^{[Fe/H]}$}
&\multicolumn{1}{p{1cm}}{\centering $P$}\\
\hline

4733677211487164160 & 47.885 & 47.885 & 1.779 & -1.770 & 0.498 & 19.243 & 67.906 & 5.324 & -1.309 & 0.211 & 0.604 \\ 
4733699790131324544 & 48.065 & 48.065 & 1.419 & -2.153 & 0.278 & 19.871 & 67.236 & 7.805 & -1.480 & 0.380 & 0.826 \\ 
4733781635030255232 & 48.139 & 48.139 & 1.519 & -1.886 & 0.478 & 19.408 & 76.810 & 2.578 & -1.362 & 0.114 & 0.930 \\ 
4733699893210538112 & 48.061 & 48.061 & 1.690 & -2.024 & 0.693 & 16.740 & 69.496 & 0.694 & -1.088 & 0.021 & 0.970 \\ 
4733781665091557760 & 48.149 & 48.149 & 1.499 & -1.817 & 0.540 & 18.339 & 74.026 & 1.240 & -1.167 & 0.065 & 0.852 \\ 
4733795585081224320 & 48.244 & 48.244 & 1.476 & -1.685 & 0.322 & 19.830 & 71.751 & 7.563 & -1.751 & 0.250 & 0.643 \\ 
4733711158908171392 & 48.726 & 48.726 & 1.830 & -2.243 & 0.331 & 19.125 & 70.250 & 3.164 & -1.288 & 0.182 & 0.914 \\ 
4733782283566915584 & 48.269 & 48.269 & 1.622 & -1.974 & 0.272 & 19.887 & 76.777 & 6.650 & -1.466 & 0.246 & 0.858 \\ 
4726956244798037120 & 44.930 & 44.930 & 2.269 & -2.879 & 0.293 & 19.201 & 35.518 & 3.020 & -1.052 & 0.179 & 0.953 \\ 
4733700335591309824 & 47.942 & 47.942 & 1.675 & -2.272 & 0.504 & 19.059 & 69.332 & 1.498 & -1.335 & 0.079 & 0.966 \\ 

\hline   
\end{tabular}
\tablefoot{The columns from left to right correspond to the following: the Gaia DR3 ID, current positions on the sky, proper motions, colors, magnitudes, line-of-sight velocities, error on the velocities, $S^5$ metallicity, error on the metallicity, and the membership probability of each star to belong the cluster.}
\end{table*}

\begin{table*}[h!]
\small
\caption{List of potential members of NGC~1904. Only 10 members are shown here while the full list is available at the CDS.}             
\label{table:ngc1904}      
\centering          
\renewcommand{\arraystretch}{1.25} 
\begin{tabular}{@{}l@{ }c@{ }c@{ }c@{ }c@{ }c@{ }c@{ }c@{ }c@{ }c@{ }c@{ }c@{ }c@{ }c@{ } c@{ }
}     
\hline 
\multicolumn{1}{p{2.5cm}}{\centering Gaia ID \\ (Gaia DR3)}
&\multicolumn{1}{p{0.9cm}}{\centering RA \\(ICRS)}
&\multicolumn{1}{p{0.9cm}}{\centering Dec \\(ICRS)}
&\multicolumn{1}{p{1.2cm}}{\centering $\mu_{\alpha}$ \\ (mas/ yr)}
&\multicolumn{1}{p{1.2cm}}{\centering $\mu_{\delta}$ \\ (mas/ yr)}
&\multicolumn{1}{p{1.4cm}}{\centering $(BP-RP)_0$ (mag)}
&\multicolumn{1}{p{0.75cm}}{\centering $G_0$ (mag)}
&\multicolumn{1}{p{0.75cm}}{\centering $v_{rad}$ \\ (km/s) }
&\multicolumn{1}{p{1cm}}{\centering $\sigma_{meas}^{v}$ \\ (km/s)}
&\multicolumn{1}{c
}{\centering [Fe/H]}
&\multicolumn{1}{p{1cm}}{\centering $\sigma_{meas}^{[Fe/H]}$}
&\multicolumn{1}{p{1cm}}{\centering $P$}\\
\hline

2957895907481574400 & 81.104 & 81.104 & 2.445 & -1.127 & 0.272 & 19.816 & 205.160 & 6.007 & -2.508 & 0.535 & 0.775 \\ 
2957939651729098752 & 81.030 & 81.030 & 2.574 & -1.574 & 0.693 & 15.210 & 207.580 & 0.664 & -1.370 & 0.012 & 0.988 \\ 
2957989400330136832 & 80.963 & 80.963 & 2.341 & -1.597 & 0.345 & 19.311 & 209.795 & 2.962 & -1.738 & 0.176 & 0.980 \\ 
2958540702332043264 & 81.865 & 81.865 & 2.365 & -1.455 & 0.495 & 19.305 & 207.331 & 5.594 & -1.696 & 0.465 & 0.921 \\ 
2958601935683680000 & 80.723 & 80.723 & 2.457 & -1.549 & 0.434 & 19.108 & 207.234 & 3.453 & -1.607 & 0.182 & 0.987 \\ 
2957934738286530688 & 81.031 & 81.031 & 2.441 & -1.637 & 0.512 & 18.163 & 205.437 & 1.150 & -1.520 & 0.063 & 0.995 \\ 
2957939686088825472 & 81.039 & 81.039 & 2.508 & -1.536 & 1.144 & 14.232 & 210.836 & 0.661 & -1.106 & 0.007 & 0.873 \\ 
2957944943128920064 & 81.233 & 81.233 & 2.768 & -1.600 & 0.273 & 19.685 & 206.844 & 6.250 & -2.362 & 0.268 & 0.894 \\ 
2957989469049599616 & 80.928 & 80.928 & 2.338 & -1.653 & 0.275 & 20.065 & 219.664 & 16.169 & -2.329 & 0.397 & 0.837 \\ 
2957945802121651072 & 81.129 & 81.129 & 2.422 & -1.571 & 113.880 & 14.880 & 203.532 & 0.667 & -1.653 & 0.013 & 0.997 \\ 

\hline   
\end{tabular}
\tablefoot{The columns from left to right correspond to the following: the Gaia DR3 ID, current positions on the sky, proper motions, colors, magnitudes, line-of-sight velocities, error on the velocities, $S^5$ metallicity, error on the metallicity, and the membership probability of each star to belong the cluster.}
\end{table*}

\end{document}